\documentclass[twocolumn,english,prd,nofootinbib]{revtex4-1}
\usepackage{lmodern}

\usepackage[T1]{fontenc}
\usepackage{color}
\usepackage{babel}
\usepackage{verbatim}
\usepackage{mathrsfs}
\usepackage{enumitem}
\usepackage{amsmath}
\usepackage{amsthm}
\usepackage{amssymb}
\usepackage{graphicx}
\usepackage[unicode=true,pdfusetitle,
 bookmarks=true,bookmarksnumbered=false,bookmarksopen=false,
 breaklinks=false,pdfborder={0 0 1},backref=false,colorlinks=true]
 {hyperref}
\hypersetup{
 pdfborderstyle=,linkcolor=blue,citecolor=blue,urlcolor={}}

\makeatletter

\theoremstyle{plain}
\newtheorem{thm}{\protect\theoremname}
\theoremstyle{plain}
\newtheorem{prop}[thm]{\protect\propositionname}

\@ifundefined{showcaptionsetup}{}{%
 \PassOptionsToPackage{caption=false}{subfig}}
\usepackage{subfig}
\makeatother

\providecommand{\propositionname}{Proposition}
\providecommand{\theoremname}{Theorem}

\begin{document}
\title{Static compact objects in Einstein-Cartan theory}
\author{Paulo Luz}
\email{paulo.luz@ist.utl.pt}

\affiliation{Center for Mathematical Analysis, Geometry and Dynamical Systems,
Instituto Superior T\'{e}cnico -- IST, Universidade de Lisboa --
UL, Avenida Rovisco Pais 1, 1049 Lisboa, Portugal}
\affiliation{Centro de Astrof\'{\i}sica e Gravita\c{c}\~{a}o - CENTRA, Departamento
de F\'{\i}sica, Instituto Superior T\'{e}cnico - IST, Universidade
de Lisboa - UL, Av. Rovisco Pais 1, 1049-001 Lisboa, Portugal}
\affiliation{Centro de Matem\'{a}tica, Universidade do Minho, Campus de Gualtar,
4710-057 Braga, Portugal}
\author{Sante Carloni}
\email{sante.carloni@gmail.com}

\affiliation{Centro de Astrof\'{\i}sica e Gravita\c{c}\~{a}o - CENTRA, Departamento
de F\'{\i}sica, Instituto Superior T\'{e}cnico - IST, Universidade
de Lisboa - UL, Av. Rovisco Pais 1, 1049-001 Lisboa, Portugal}
\begin{abstract}
We generalize the Tolman-Oppenheimer-Volkoff equations for space-times
endowed with a Weyssenhoff like torsion field in the Einstein-Cartan
theory. The new set of structure equations clearly show how the presence
of torsion affects the geometry of the space-time. We obtain new exact
solutions for compact objects with non-null intrinsic spin surrounded by vacuum,
explore their properties and discuss how these solutions should be
smoothly matched to an exterior space-time. We study how the intrinsic spin of matter changes
the Buchdahl limit for the maximum compactness of stars. Moreover,
under rather generic conditions, we prove that in the context of a
Weyssenhoff like torsion, no static, spherically symmetric compact
objects supported only by the intrinsic spin can exist. We also provide some
algorithms to generate new solutions.
\end{abstract}
\maketitle

\section{Introduction}

Compact objects, in particular neutron stars, represent one of richest
environments to probe fundamental physics due their extreme gravitational
fields, densities and the state of the matter that composes them,
especially, at the core. The recent detection of gravitational waves
due to the coalescing of two orbiting neutron stars \citep{LIGO}
opened a new window to study their tidal deformations, allowing
the study of the properties of the matter fields that compose this
kind of objects. Nonetheless, the usage of neutron stars as a physics
laboratory is only possible if we have a deep knowledge of their properties.
In particular, it is important to understand how the intrinsic spin\footnote{We should remark that here, and in the following, the word ``spin'' will be used exclusively to represent the quantum spin of the particles that source the gravitational field equations. In no case the word spin will be associated to any form of rotation of the compact objects we will analyze.} of the fermionic
matter particles affects the behavior of such bodies.

In an astrophysical context, the effects of spin were first considered
when Chandrasekhar \citep{Chandrasekhar} established the maximum
mass an ideal white dwarf could hold due to the electron degeneracy
pressure, before it underwent continuous gravitational collapse (see
also Ref.~\citep{Durisen} for the rotating case). In the subsequent
years, similar limits relying on the Pauli exclusion principle were
proposed for other types of compact objects, namely neutron stars,
showing that the spin of matter particles markedly influences astrophysical
objects (cf. e.g. Ref.~\citep{Hawking_Israel}). Nevertheless, the
way in which the presence of intrinsic spin affects the properties
of astrophysical bodies, remains largely unknown.

In a affine theory of gravity, the gravitational field is represented
by the geometry of the space-time which is, in turn, is determined by
the energy and momentum of the matter fields. Mathematically, all
classical matter properties are described by an energy-momentum tensor
that acts as a source in the field equations.
Since spin can be considered as an intrinsic angular momentum of the
matter particles, one would expect that this property could also
be encoded in an energy-momentum tensor. However, in
the theory of General Relativity (GR) it appears immediately clear
that there is no obvious way to introduce the spin in a way that is
consistent with the conservation laws for the total angular momentum.
A way around this problem is to endow the space-time with additional
geometrical structure, providing extra degrees of freedom to model
spin and its relation with the gravitational field. This is the fundamental
idea behind the so-called Einstein-Cartan-Sciama-Kible (ECSK) theory
of gravity. In this theory the connection is not imposed be symmetric
so that, the anti-symmetric part of the connection defines an extra
tensor field: torsion. In this way, it is possible to impose a local
Poincar\'{e} gauge symmetry on the tangent space of each point of
the manifold such that the matter intrinsic spin can be related with
the torsion tensor field. Indeed, theories of gravity with a non-symmetric
connection (generically called Einstein-Cartan theories) were proposed
even before the discovery of spin. Sciama and Kibble \citep{Sciama,Kibble}
introduced the idea of connecting the torsion tensor with the matter
intrinsic spin, paving the way to a geometrized treatment of spin.

Early works on the ECSK theory focused on the effects of spin on the
evolution of gravitational collapse and the possibility of avoidance
of singularities \citep{Trautman,Stewart,Kop,Hehl2,Hehl3}. Only by the
end of the decade, solutions for spherically symmetric space-times
were found \citep{Singh,DemPros}. The solutions in Ref.~\citep{Singh}
were obtained by directly solving the field equations for the ECSK theory.
Such approach, though, leads to great difficulties in searching for
exact solutions. In this article we will adopt a different method
and consider the formalism provided by the 1+1+2 space-time decomposition
\citep{Elst,Clarkson_1,Betschart,Clarkson_2}. Covariant space-time
decomposition approaches were initially devised as a powerful tool
to explore the properties of cosmological models and their perturbations
(see e.g. \citep{Elst,Ehlers,Ellis_Maartens,Ellis:1998ct}) and only
recently they have been employed to deal with space-times of astrophysical
interest. In ref.~\citep{Sante1,Sante2}, this approach was used
to construct - in the context of GR - a covariant version of the Tolman-Oppenheimer-Volkoff
equations. The new equations allowed to pinpoint the mathematical
nature of the problem of determining interior solutions for compact
objects and, for instance, the treatment of stars with anisotropic
pressure. Moreover, in the covariant language it was possible to define
algorithms to generate a number of new exact solutions, and to easily
obtain general theorems (like the ones \citep{Boonserm1,Boonserm2}
in GR) which link apparently unrelated solutions.

In this article, we aim to study static compact objects in the context
of the ECSK theory, in particular, study how the presence of spin
affects the possible solutions. Moreover, we will also examine how
the boundary conditions imposed by the smooth junction of two space-times,
with possible non-null torsion, constraint the solutions.

The article is organized as follows: in Section \ref{sec:The_1p1p2_decomposition}
we define the 1+1+2 formalism and consider the decomposition of some
tensorial quantities; in Section \ref{sec:Structure_equations} we
describe the setup that we propose to study and provide the structure
equations; in Section \ref{sec:Generalized_TOV_equation} we derive
the Tolman-Oppenheimer-Volkoff (TOV) equations for static, locally
rotationally symmetric space-times of class I and II in the presence
of a non-null torsion field; in Section \ref{sec:Junction_Conditions}
we generalize the conditions for the smooth junction of two space-times
with general torsion tensor fields and apply the results to the particular
considered setup; in Section \ref{sec:Exact_solutions} exact solutions
are derived and studied; in Section \ref{sec:Generating_theorems}
we provide a set of algorithms to generate new exact solutions from
previously known ones and in Section \ref{sec:Conclusions} we we
summarize the results and conclude.

In this article we shall assume the metric signature $\left(-+++\right)$
and work in the geometrized units system where $G=c=1$.

\section{The 1+1+2 decomposition\label{sec:The_1p1p2_decomposition}}

Consider a Lorentzian manifold of dimension 4 and a congruence of
time-like curves with tangent vector $u$. Without loss of generality
we can foliate the manifold in 3-hypersurfaces, $V$, orthogonal at
each point to the curves of the congruence, such that all quantities
are defined by their behavior along the direction of $u$ and in $V$.
This procedure is usually called ``1+3 space-time decomposition''.
Such decomposition of the space-time manifold relies on the existence
of a projector to the hypersurface $V$ which can be naturally defined
as 
\begin{equation}
h_{\alpha\beta}=g_{\alpha\beta}+u_{\alpha}u_{\beta}\,,\label{eq:projector_h_definition}
\end{equation}
where $g_{\alpha\beta}$ represents the space-time metric and $u_{\alpha}u^{\alpha}=-1$. the projector $h_{\alpha\beta}$ has the following properties 
\begin{equation}
\begin{aligned}h_{\alpha\beta} & =h_{\beta\alpha}\,, &  &  & h_{\alpha\beta}h^{\beta\gamma} & =h_{\alpha}^{\gamma}\,,\\
h_{\alpha\beta}u^{\alpha} & =0\,, &  &  & h_{\alpha}^{\alpha} & =3\,.
\end{aligned}
\end{equation}

The 1+1+2 decomposition \citep{Elst,Clarkson_1,Betschart,Clarkson_2}
builds from the 1+3 decomposition by defining a congruence of spatial
curves with tangent vector field $e$ such that any quantity defined
in the sub-manifold $V$ is defined by its behavior along $e$ and
in the 2-surfaces $W$. We shall refer to $W$ as ``the sheet''.
As before, we can then define a projector onto $W$ by 
\begin{equation}
N_{\alpha\beta}=h_{\alpha\beta}-e_{\alpha}e_{\beta}\,,\label{eq:projector_N_definition}
\end{equation}
where $e_{\alpha}e^{\alpha}=1$, and such that

\begin{equation}
\begin{aligned}N_{\alpha\beta} & =N_{\beta\alpha}\,, &  &  & N_{\alpha\beta}N^{\beta\gamma} & =N_{\alpha}^{\gamma}\,,\\
N_{\alpha\beta}u^{\alpha} & =N_{\alpha\beta}e^{\alpha}=0\,, &  &  & N_{\alpha}^{\alpha} & =2\,.
\end{aligned}
\label{eq:projector_N_properties}
\end{equation}

It is useful to introduce the following tensors 
\begin{equation}
\begin{aligned}\varepsilon_{\alpha\beta\gamma} & =\varepsilon_{\alpha\beta\gamma\sigma}u^{\sigma}\,,\\
\varepsilon_{\alpha\beta} & =\varepsilon_{\alpha\beta\gamma}e^{\gamma}\,,
\end{aligned}
\label{eq:volume_forms}
\end{equation}
derived from the covariant Levi-Civita tensor $\varepsilon_{\alpha\beta\gamma\sigma}$,
with the following properties 
\begin{equation}
\begin{aligned}\varepsilon_{\alpha\beta\gamma} & =\varepsilon_{\left[\alpha\beta\gamma\right]}\,, &  &  & \varepsilon_{\alpha\beta} & =\varepsilon_{\left[\alpha\beta\right]}\,,\\
\varepsilon_{\alpha\beta\gamma}u^{\gamma} & =0\,, &  &  & \varepsilon_{\alpha\beta}u^{\alpha} & =\varepsilon_{\alpha\beta}e^{\alpha}=0\,,\\
\varepsilon_{\alpha\beta\gamma}\varepsilon^{\mu\nu\gamma} & =h_{\alpha}^{\mu}h_{\beta}^{\nu}-h_{\beta}^{\mu}h_{\alpha}^{\nu}\,, &  &  & \varepsilon_{\alpha}{}^{\gamma}\varepsilon_{\beta\gamma} & =N_{\alpha\beta}\,,\\
\varepsilon_{\mu\nu\alpha}\varepsilon^{\mu\nu\beta} & =2h_{\alpha}^{\beta}\,, &  &  & \varepsilon_{\alpha\beta\gamma} & =e_{\alpha}\varepsilon_{\beta\gamma}-e_{\beta}\varepsilon_{\alpha\gamma}+\\
 &  &  &  &  & \,\,\,\,+e_{\gamma}\varepsilon_{\alpha\beta}\,.
\end{aligned}
\end{equation}

Using the results in appendix \ref{Appendix: Covariant_quantities},
the covariant derivatives of the tangent vectors $u$ and $e$ can
be written as

\begin{equation}
\begin{aligned}\delta_{\alpha}u_{\beta}=N_{\alpha}^{\sigma}N_{\beta}^{\gamma}\nabla_{\sigma}u_{\gamma} & =N_{\alpha\beta}\left(\frac{1}{3}\theta-\frac{1}{2}\Sigma\right)+\Sigma_{\alpha\beta}+\varepsilon_{\alpha\beta}\Omega\,,\\
D_{\alpha}u_{\beta}=h_{\alpha}^{\sigma}h_{\beta}^{\gamma}\nabla_{\sigma}u_{\gamma} & =\delta_{\alpha}u_{\beta}+\left(\frac{1}{3}\theta+\Sigma\right)e_{\alpha}e_{\beta}\\
 & +2\Sigma_{(\alpha}e_{\beta)}-\varepsilon_{\alpha\lambda}\Omega^{\lambda}e_{\beta}+e_{\alpha}\varepsilon_{\beta\lambda}\Omega^{\lambda}\,,\\
\nabla_{\alpha}u_{\beta} & =D_{\alpha}u_{\beta}-u_{\alpha}\left(\mathcal{A}e_{\beta}+\mathcal{A}_{\beta}\right)\,,
\end{aligned}
\end{equation}
and 
\begin{equation}
\begin{aligned}\delta_{\alpha}e_{\beta}= & \frac{1}{2}N_{\alpha\beta}\phi+\zeta_{\alpha\beta}+\varepsilon_{\alpha\beta}\xi\,,\\
D_{\alpha}e_{\beta}= & \delta_{\alpha}e_{\beta}+e_{\alpha}a_{\beta}\,,\\
\nabla_{\alpha}e_{\beta}= & D_{\alpha}e_{\beta}-u_{\alpha}\alpha_{\beta}-\mathcal{A}u_{\alpha}u_{\beta}+\left(\frac{1}{3}\theta+\Sigma\right)e_{\alpha}u_{\beta}+\\
 & +\left(\Sigma_{\alpha}-\varepsilon_{\alpha\sigma}\Omega^{\sigma}\right)u_{\beta}\,.
\end{aligned}
\label{eq:Cov_vector_e}
\end{equation}

We shall also need to find the various contributions along $u$, $e$
and on $W$ of the energy-momentum tensor $\mathcal{T}_{\alpha\beta}$.
At this point we shall not assume $\mathcal{T}_{\alpha\beta}$ to
have any symmetry. Hence, 
\begin{equation}
\begin{aligned}\mathcal{T}_{\alpha\beta} & =\mu\,u_{\alpha}u_{\beta}+p\,h_{\alpha\beta}+q_{1\alpha}u_{\beta}+u_{\alpha}q_{2\beta}+\pi_{\alpha\beta}+m_{\alpha\beta}\\
 & =\mu\,u_{\alpha}u_{\beta}+Q_{1\alpha}u_{\beta}+u_{\alpha}Q_{2\beta}+Q_{1}\,e_{\alpha}u_{\beta}+\\
 & +Q_{2}\,u_{\alpha}e_{\beta}+p_{r}\,e_{\alpha}e_{\beta}+\Pi_{1\alpha}e_{\beta}+e_{\alpha}\Pi_{2\beta}+\\
 & +p_{\perp}\,N_{\alpha\beta}+\Pi_{\alpha\beta}+\varepsilon_{\alpha\beta}\mathbb{M}\,,
\end{aligned}
\label{eq:Energy_momentum_tensor_decomposition_general}
\end{equation}
with 
\begin{equation}
\begin{aligned}q_{1\alpha} & =-h_{\alpha}^{\sigma}u^{\gamma}\mathcal{T}_{\sigma\gamma}\,, &  &  & \mu & =u^{\sigma}u^{\gamma}\mathcal{T}_{\sigma\gamma}\,,\\
q_{2\alpha} & =-u^{\sigma}h_{\alpha}^{\gamma}\mathcal{T}_{\sigma\gamma}\,, &  &  & p & =\frac{1}{3}h^{\alpha\beta}\mathcal{T}_{\alpha\beta}\,,\\
Q_{1\alpha} & =-N_{\alpha}^{\sigma}u^{\gamma}\mathcal{T}_{\sigma\gamma}\,, &  &  & p_{r} & =p+\Pi=e^{\sigma}e^{\gamma}\mathcal{T}_{\sigma\gamma}\,,\\
Q_{2\alpha} & =-u^{\sigma}N_{\alpha}^{\gamma}\mathcal{T}_{\sigma\gamma}\,, &  &  & p_{\perp} & =p-\frac{1}{2}\Pi=\frac{1}{2}N^{\sigma\gamma}\mathcal{T}_{\sigma\gamma}\,,\\
\Pi_{1\alpha} & =N_{\alpha}^{\sigma}e^{\gamma}\mathcal{T}_{\sigma\gamma}\,, &  &  & Q_{1} & =-e^{\sigma}u^{\gamma}\mathcal{T}_{\sigma\gamma}\,,\\
\Pi_{2\alpha} & =e^{\sigma}N_{\alpha}^{\gamma}\mathcal{T}_{\sigma\gamma}\,, &  &  & Q_{2} & =-u^{\sigma}e^{\gamma}\mathcal{T}_{\sigma\gamma}\,,\\
\pi_{\alpha\beta} & =h_{\left\langle \alpha\right.}^{\sigma}h_{\left.\beta\right\rangle }{}^{\gamma}\mathcal{T}_{\sigma\gamma}\,, &  &  & \Pi & =\frac{1}{3}\mathcal{T}_{\alpha\beta}\left(2e^{\alpha}e^{\beta}-N^{\alpha\beta}\right)\,,\\
m_{\alpha\beta} & =h_{\left[\alpha\right.}^{\sigma}h_{\left.\beta\right]}{}^{\gamma}\mathcal{T}_{\sigma\gamma}\,, &  &  & \mathbb{M} & =\frac{1}{2}\varepsilon^{\mu\nu}\mathcal{T}_{\mu\nu}\,,\\
\Pi_{\alpha\beta} & =\mathcal{T}_{\left\{ \alpha\beta\right\} }\,,
\end{aligned}
\label{eq:Energy_momentum_tensor_decomposition_quantities}
\end{equation}
where the angular and curly parentheses notation
is defined in Eq.~\eqref{eqA:curly_notation_definition}. Moreover,
the following relations are useful
\begin{equation}
\begin{aligned}q_{1,2\alpha} & =Q_{1,2\alpha}+Q_{1,2}\,e_{\alpha}\,,\\
\pi_{\alpha\beta} & =\Pi_{\alpha\beta}+\Pi\left(e_{\alpha}e_{\beta}-\frac{1}{2}N_{\alpha\beta}\right)+\\
 & +\Pi_{1\left(\alpha\right.}e_{\left.\beta\right)}+\Pi_{2\left(\alpha\right.}e_{\left.\beta\right)}\,.
\end{aligned}
\end{equation}

In this paper we will assume that the space-time is endowed with
a linear, metric compatible connection $C_{\alpha\beta}{}^{\gamma}$.
Such connection is characterized by the metric connection - the Christoffel
symbols - and the torsion tensor field 
\[
S_{\alpha\beta}\,^{\gamma}=C_{\left[\alpha\beta\right]}^{\gamma}\,.
\]
Using Eq.~\eqref{eq:projector_h_definition} we can write the torsion
tensor field as 
\begin{equation}
S_{\alpha\beta\gamma}=\varepsilon_{\alpha\beta}\,^{\mu}\bar{S}_{\mu\gamma}+W_{[\alpha|\gamma}u_{|\beta]}+S_{\alpha\beta}u_{\gamma}+u_{[\alpha}X_{\beta]}u_{\gamma}\,,\label{eq:timelike_torsion_decomposition}
\end{equation}
with 
\begin{equation}
\begin{aligned}\bar{S}_{\alpha\beta} & =\frac{1}{2}\varepsilon_{\alpha\mu\nu}h_{\beta}^{\sigma}S^{\mu\nu}{}_{\sigma}\,,\\
W_{\alpha\beta} & =2u^{\mu}h_{\alpha}^{\nu}h_{\beta}^{\sigma}S_{\mu\nu\sigma}\,,\\
S_{\alpha\beta} & =-h_{\alpha}^{\mu}h_{\beta}^{\nu}u^{\sigma}S_{\mu\nu\sigma}\,,\\
X_{\alpha} & =2u^{\mu}h_{\alpha}^{\nu}u^{\sigma}S_{\mu\nu\sigma}\,.
\end{aligned}
\label{eq:timelike_torsion_decomposition_components_def}
\end{equation}
Notice that the tensors defined in Eq.~\eqref{eq:timelike_torsion_decomposition_components_def}
are orthogonal to the tangent vector $u$.

Now, from the definition of the Riemann tensor, $R_{\alpha\beta\gamma}{}^{\delta}$:
\begin{equation}
R_{\alpha\beta\gamma}{}^{\delta}w_{\delta}=\nabla_{\alpha}\nabla_{\beta}w_{\gamma}-\nabla_{\beta}\nabla_{\alpha}w_{\gamma}+2S_{\alpha\beta}{}^{\delta}\nabla_{\delta}w_{\gamma}\,,\label{eq:Riemann_tensor_definition}
\end{equation}
where $w_{\gamma}$ is an arbitrary 1-form; in the case of a Lorentzian
manifold with non-null torsion, the Riemann curvature tensor does
not possess the same symmetries as the torsion free case. This is
discussed very thoroughly for instance in Ref.~\citep{Jensen} to
which we redirect the reader for further details. In theories with
torsion the Riemann tensor has the following properties: 
\begin{equation}
\begin{aligned}R_{\alpha\beta\gamma\delta} & =-R_{\beta\alpha\gamma\delta}\,,\\
R_{\alpha\beta\gamma\delta} & =-R_{\alpha\beta\delta\gamma}\,,\\
R_{\left[\alpha\beta\gamma\right]}{}^{\delta} & =-2\nabla_{\left[\alpha\right.}S_{\left.\beta\gamma\right]}{}^{\delta}+4S_{\left[\alpha\beta\right|}{}^{\rho}S_{\left|\gamma\right]\rho}{}^{\delta}\,,
\end{aligned}
\label{eq:Riemann_tensor_properties}
\end{equation}
and the modified second Bianchi identity 
\begin{equation}
\nabla_{\left[\alpha\right.}R_{\left.\beta\gamma\right]\delta}{}^{\rho}=Q_{\alpha\beta\gamma\delta}{}^{\rho}\,,\label{eq:second_Bianchi_identity}
\end{equation}
where 
\begin{equation}
Q_{\alpha\beta\gamma\delta\rho}=2S_{\left[\alpha\beta\right|}{}^{\sigma}R_{\left|\gamma\right]\sigma\delta\rho}\,.\label{eq:B_tensor_definition}
\end{equation}

The previous properties of the Riemann tensor are completely general,
in particular they are valid for space-times of any dimension. Let
us now consider the particular case of a space-time of dimension 4
with torsion. In this case, $R_{\alpha\beta\gamma\delta}$ can be
written as the following sum 
\begin{equation}
R_{\alpha\beta\gamma\delta}=C_{\alpha\beta\gamma\delta}+R_{\alpha\left[\gamma\right.}g_{\left.\delta\right]\beta}-R_{\beta\left[\gamma\right.}g_{\left.\delta\right]\alpha}-\frac{1}{3}R\,g_{\alpha\left[\gamma\right.}g_{\left.\delta\right]\beta}\,,\label{eq:Weyl_tensor_definition}
\end{equation}
where $C_{\alpha\beta\gamma\delta}$ represents the Weyl tensor, $R_{\alpha\beta}=R_{\alpha\mu\beta}{}^{\mu}$
the Ricci tensor and $R$ the Ricci scalar. In precence of torsion,
the Weyl tensor is still defined as the trace-free part of the curvature
tensor, but it does not retain all the other usual symmetries: 
\begin{equation}
\begin{aligned}C_{\alpha\beta\gamma\delta} & =-C_{\beta\alpha\gamma\delta}\,,\\
C_{\alpha\beta\gamma\delta} & =-C_{\alpha\beta\delta\gamma}\,,\\
C_{\left[\alpha\beta\gamma\right]\delta} & =R_{\left[\alpha\beta\gamma\right]\delta}+R_{\left[\alpha\beta\right.}g_{\left.\gamma\right]\delta}\,.
\end{aligned}
\label{eq:Weyl_tensor_properties}
\end{equation}
It will be useful for our purposes to give the relation between the
derivative of the Weyl tensor and the Riemann tensor. The generalization
of the formula given in Ref.~\citep{Trumper} (see also Ref.~\citep{Ellis_Maartens})
is 
\begin{equation}
\begin{aligned}\nabla_{\alpha}C^{\gamma\delta\beta\alpha} & =\frac{1}{4}\varepsilon^{\mu\nu\lambda\beta}Q_{\mu\nu\lambda\sigma\rho}\varepsilon^{\sigma\rho\gamma\delta}+\\
 & \;+\frac{3}{2}g^{\beta\left[\delta\right.}Q^{\left.\gamma\right]\mu\nu}{}_{\mu\nu}+\nabla^{\left[\delta\right.}R^{\left.\gamma\right]\beta}-\frac{1}{6}g^{\beta\left[\gamma\right.}\nabla^{\left.\delta\right]}R\,,
\end{aligned}
\end{equation}
where $Q_{\alpha\beta\gamma\delta\rho}$ is given by Eq.~\eqref{eq:B_tensor_definition}.

In what follows, we will need to find the 1+1+2 decomposition of the
Weyl tensor. We start by decomposing $C_{\alpha\beta\gamma\delta}$
into its components along $u$ and $V$ as 
\begin{equation}
\begin{aligned}C_{\alpha\beta\gamma\delta} & =-\varepsilon_{\alpha\beta\mu}\varepsilon_{\gamma\delta\nu}E^{\nu\mu}-2u_{\alpha}E_{\beta\left[\gamma\right.}u_{\left.\delta\right]}+2u_{\beta}E_{\alpha\left[\gamma\right.}u_{\left.\delta\right]}-\\
 & -2\varepsilon_{\alpha\beta\mu}H^{\mu}{}_{\left[\gamma\right.}u_{\left.\delta\right]}-2\varepsilon_{\mu\gamma\delta}\bar{H}^{\mu}{}_{\left[\alpha\right.}u_{\left.\beta\right]}\,,
\end{aligned}
\label{eq:Weyl_tensor_decomposition}
\end{equation}
where 
\begin{align}
E_{\alpha\beta} & =C_{\alpha\mu\beta\nu}u^{\mu}u^{\nu}\,,\\
H_{\alpha\beta} & =\frac{1}{2}\varepsilon_{\alpha}{}^{\mu\nu}C_{\mu\nu\beta\delta}u^{\delta}\,,\\
\bar{H}_{\alpha\beta} & =\frac{1}{2}\varepsilon_{\alpha}{}^{\mu\nu}C_{\beta\delta\mu\nu}u^{\delta}\,,\label{eq:Weyl_tensor_magnetic2}
\end{align}
are the ``electric'' part and ``magnetic'' parts of the Weyl tensor,
respectively. Eq. \eqref{eq:Weyl_tensor_decomposition} generalizes
the results in Ref.~\citep{Ellis_1971} to the case of non-null torsion.
Note, however, that, differently from the torsionless case, there
are two different tensor quantities associated to the magnetic part
of the Weyl tensor: $H_{\alpha\beta}$ and $\bar{H}_{\alpha\beta}$.

From the results in Eq.~\eqref{eq:Weyl_tensor_properties}, we see
that in the presence of torsion the tensors $E_{\alpha\beta}$, $H_{\alpha\beta}$
and $\bar{H}_{\alpha\beta}$ have the properties: 
\begin{equation}
\begin{aligned}H_{\alpha\beta} & =h_{\alpha}^{\mu}h_{\beta}^{\nu}H_{\mu\nu}\,, &  &  & H_{\alpha\beta} & =H_{\left(\alpha\beta\right)}\,,\\
\bar{H}_{\alpha\beta} & =h_{\alpha}^{\mu}h_{\beta}^{\nu}\bar{H}_{\mu\nu}\,, &  &  & \bar{H}_{\alpha\beta} & =\bar{H}_{\left(\alpha\beta\right)}\,,\\
E_{\alpha\beta} & =h_{\alpha}^{\mu}h_{\beta}^{\nu}E_{\mu\nu}\,, &  &  & E^{\alpha}{}_{\alpha} & =0\,.
\end{aligned}
\end{equation}
Therefore, $E_{\alpha\beta}$, may not be a symmetric tensor and $H_{\alpha\beta}$
and $\bar{H}_{\alpha\beta}$ do not have to be trace-free. On the
other hand, due to the properties of the Levi-Civita tensor, even
in the presence of torsion, the magnetic parts of the Weyl tensor
are symmetric under the exchange of indexes. These properties allow
us to decompose the tensors $E_{\alpha\beta}$, $H_{\alpha\beta}$
and $\bar{H}_{\alpha\beta}$ as 
\begin{align}
E_{\alpha\beta} & =\mathcal{E}\left(e_{\alpha}e_{\beta}-\frac{1}{2}N_{\alpha\beta}\right)+\mathcal{E}_{\alpha}e_{\beta}+e_{\alpha}\bar{\mathcal{E}}_{\beta}+\mathcal{E}_{\alpha\beta}+\varepsilon_{\alpha\beta}\mathbb{E}\,,\\
H_{\alpha\beta} & =\frac{1}{2}N_{\alpha\beta}\mathrm{H}+e_{\alpha}e_{\beta}\mathcal{H}+\mathcal{H}_{\alpha}e_{\beta}+e_{\alpha}\mathcal{H}_{\beta}+\mathcal{H}_{\alpha\beta}\,,\\
\bar{H}_{\alpha\beta} & =\frac{1}{2}N_{\alpha\beta}\bar{\mathrm{H}}+e_{\alpha}e_{\beta}\bar{\mathcal{H}}+\bar{\mathcal{H}}_{\alpha}e_{\beta}+e_{\alpha}\bar{\mathcal{H}}_{\beta}+\bar{\mathcal{H}}_{\alpha\beta}\,,
\end{align}
with 
\begin{equation}
\begin{aligned}\mathcal{E} & =E_{\mu\nu}e^{\mu}e^{\nu}=-N^{\mu\nu}E_{\mu\nu}\,, &  &  & \mathcal{E}_{\alpha} & =N_{\alpha}^{\mu}e^{\nu}E_{\mu\nu}\,,\\
\mathbb{E} & =\frac{1}{2}\varepsilon^{\mu\nu}E_{\mu\nu}\,, &  &  & \bar{\mathcal{E}}_{\alpha} & =e^{\mu}N_{\alpha}^{\nu}E_{\mu\nu}\,,\\
\mathrm{H} & =N^{\mu\nu}H_{\mu\nu}\,, &  &  & \mathcal{E}_{\alpha\beta} & =E_{\left\{ \alpha\beta\right\} }\,,\\
\mathcal{H} & =e^{\mu}e^{\nu}H_{\mu\nu}\,, &  &  & \mathcal{H}_{\alpha} & =N_{\alpha}^{\mu}e^{\nu}H_{\mu\nu}\,,\\
\bar{\mathrm{H}} & =N^{\mu\nu}\bar{H}_{\mu\nu}\,, &  &  & \bar{\mathcal{H}}_{\alpha} & =N_{\alpha}^{\mu}e^{\nu}\bar{H}_{\mu\nu}\,,\\
\bar{\mathcal{H}} & =e^{\mu}e^{\nu}\bar{H}_{\mu\nu}\,, &  &  & \mathcal{H}_{\alpha\beta} & =H_{\left\{ \alpha\beta\right\} }\,,\\
 &  &  &  & \bar{\mathcal{H}}_{\alpha\beta} & =\bar{H}_{\left\{ \alpha\beta\right\} }\,.
\end{aligned}
\label{eq:Weyl_tensor_components_definition}
\end{equation}
where the curly parentheses notation is defined in Eq.~\eqref{eqA:curly_notation_definition}.

\section{Decomposition of the field equations\label{sec:Structure_equations}}

We are now in position to apply that framework to study solutions
of the Einstein-Cartan theory, characterized by the following field
equations

\begin{align}
R_{\alpha\beta}-\frac{1}{2}g_{\alpha\beta}R & =8\pi\mathcal{T}_{\alpha\beta}\,,\label{eq:Weissenhoff_efe1}\\
S^{\alpha\beta\gamma}+2g^{\gamma[\alpha}S^{\beta]}{}_{\mu}{}^{\mu} & =-8\pi\Delta^{\alpha\beta\gamma}\,,\label{eq:Weissenhoff_efe2}
\end{align}
where $\mathcal{T}_{\alpha\beta}$ represents the canonical energy-momentum
tensor and $\Delta^{\alpha\beta\mu}$ the intrinsic hypermomentum,
found by varying, independently, the Einstein-Hilbert action with
respect to the metric and to the connection. We assume a null cosmological
constant.

From the second Bianchi identity \eqref{eq:second_Bianchi_identity}
and the field equations \eqref{eq:Weissenhoff_efe1} and \eqref{eq:Weissenhoff_efe2}
we find the conservation laws for the canonical energy-momentum tensor:
\begin{equation}
\nabla_{\beta}\mathcal{T}_{\alpha}{}^{\beta}=2S_{\alpha\mu\nu}\mathcal{T}^{\nu\mu}+\frac{1}{8\pi}\left(S_{\alpha\mu}{}^{\mu}R-S^{\mu\nu\sigma}R_{\alpha\sigma\mu\nu}\right)\,.
\end{equation}
Using Eq.~\eqref{eq:Weissenhoff_efe2} we can introduce the intrinsic
hypermomentum tensor in the above conservation laws and recover the
result in Ref.~\citep{Hehl1} (see also Ref.~\citep{Puetzfeld,Yasskin}
for similar results derived in a more general context).

We will assume that the source for the above field equations is an
uncharged Weyssenhoff fluid \citep{Weyssenhoff}. The Weyssenhoff
fluid provides a semi-classical description of a perfect fluid composed
of fermions, such that the fluid is characterized by its energy density,
pressure and spin density. Its canonical energy-momentum tensor is
given by 
\begin{equation}
\mathcal{T}_{\alpha\beta}=\mu u_{\alpha}u_{\beta}+p\,h_{\alpha\beta}-\left(\mathcal{A}e^{\mu}+\mathcal{A^{\mu}}\right)S_{\mu\alpha}u_{\beta}\,,\label{eq:Weyssenhoff_energy_momentum_tensor}
\end{equation}
where $\mu$ and $p$ represent the energy density and pressure of
the fluid, respectively.

Following Refs.~\citep{Smalley,obukhov1}, the hypermomentum tensor
for the Weyssenhoff spin fluid can be written, in our conventions,
as 
\begin{equation}
\Delta^{\alpha\beta\gamma}=-\frac{1}{8\pi}\Delta^{\alpha\beta}u^{\gamma}\,,
\end{equation}
where $u$ represents the proper 4-velocity of an element of volume
of the fluid and the anti-symmetric spin density tensor, $\Delta^{\alpha\beta}$,
verifies $\Delta^{\alpha\beta}u_{\beta}=0$. From Eq.~\eqref{eq:Weissenhoff_efe2}
we find, 
\begin{equation}
S^{\alpha\beta\gamma}=\Delta^{\alpha\beta}u^{\gamma}\,,\label{eq:Weyssenhoff_torsion}
\end{equation}
hence, comparing Eq.~\eqref{eq:Weyssenhoff_torsion} with Eq.~\eqref{eq:timelike_torsion_decomposition}
we see that the Weyssenhoff fluid model implies that the tensors $\bar{S}^{\alpha\beta}$,
$W^{\alpha\beta}$ and $X^{\alpha}$, Eq.~\eqref{eq:timelike_torsion_decomposition_components_def},
are null and $S^{\alpha\beta}=\Delta^{\alpha\beta}$. In this way,
the decomposition of the torsion tensor will coincide with the decomposition
of $S^{\alpha\beta}$. Taking into account that $S_{\alpha\beta}\equiv S_{\left[\alpha\beta\right]}$
we find 
\begin{equation}
S_{\alpha\beta}=\varepsilon_{\alpha\beta}\tau+2s_{\left[\alpha\right.}e_{\left.\beta\right]}\,,\label{eq:timelike_torsion_decomposition_S_tensor}
\end{equation}
with 
\begin{equation}
\begin{aligned}\tau & =\frac{1}{2}\varepsilon^{\mu\nu}S_{\mu\nu}\,,\\
s_{\alpha} & =N_{\alpha}^{\sigma}e^{\gamma}S_{\sigma\gamma}\,.
\end{aligned}
\label{eq:timelike_torsion_decomposition_S_tensor_components}
\end{equation}

\subsection{The symmetries of the problem}

We are interested in solutions of the Einstein-Cartan theory that
are static and locally rotationally symmetric (LRS). Following Ref.~\citep{Stewart_Ellis},
a space-time is said to be local rotational symmetric in a neighborhood
of a point $q$, $B\left(q\right)$, if there exists a non-discrete
sub-group $G$ of the Lorentz group in the tangent space of each $q'\in B\left(q\right)$
which leaves $u$, the curvature tensor and their derivatives (up
to third order) invariant. Assuming $G$ to be one-dimensional, we
can set at each point the vector field $e$ to have the same direction
as an axis of symmetry. Then, LRS implies that all covariantly defined
space-like vectors must have the same direction of $e$ - otherwise
they would not be invariant under $G$. Thus, the vector quantities
$\left\{ a_{\beta},\,\alpha_{\beta},\,\Sigma_{\beta},\,\Omega_{\beta}\,,A_{\beta}\right\} $
are null in such space-times. Also the shear tensors of the congruences
of curves associated with $u$ and $e$ projected onto the sheet:
$\Sigma_{\alpha\beta}$ and $\zeta_{\alpha\beta}$, must be null since,
there can not be any preferred direction at the sheet\footnote{It should be remarked here that the presence of a generic torsion
tensor field affects the definition of the kinematical quantities
\citep{Luz,Paoli,Liberati,Speziale}. See the Appendix \ref{Appendix:subsec_Physical_kinematical_quantities}
for further details. As such, in the presence of a general torsion,
LRS implies that the geometric shear vector fields $\Sigma_{g\,\alpha\beta}$
and $\zeta_{g\,\alpha\beta}$ must be null and not the quantities
$\Sigma_{\alpha\beta}=\sigma_{\left\{ \alpha\beta\right\} }$ and
$\zeta_{\alpha\beta}\equiv\delta_{\{\alpha}e_{\beta\}}$. However,
as discussed in Appendix \ref{Appendix:subsec_Physical_kinematical_quantities},
for a Weyssenhoff fluid those are equal hence, from here on out we
shall refer to $\Sigma_{\alpha\beta}$ and $\zeta_{\alpha\beta}$
as the shear tensors, onto the sheet, of the congruences associated
with $u$ and $e$, being implicit that we assume the Weyssenhoff
model.}.

From the definition of LRS space-times, the Riemann curvature tensor
must also be invariant under $G$ therefore, the vector components
of the Weyl tensor $\left\{ \mathcal{E}_{1\alpha},\,\mathcal{E}_{2\alpha},\,\left(\mathcal{H}_{,2}\right)_{\alpha}\right\} $
must also be null. Since the Riemann tensor also depends on the torsion
tensor, the latter must also be invariant under the action of $G$.
Therefore, from Eqs.~\eqref{eq:Weyssenhoff_torsion} and \eqref{eq:timelike_torsion_decomposition_S_tensor},
the tensor field $s^{\alpha}$, Eq.~\eqref{eq:timelike_torsion_decomposition_S_tensor_components},
must be null. In light of this results and taking into account Eq.~\eqref{eq:Weyssenhoff_torsion}
we also conclude that for an LRS space-time the intrinsic hypermomentum
tensor is simply given by 
\begin{equation}
\Delta_{\alpha\beta}=\varepsilon_{\alpha\beta}\delta\,,\label{eq:Weyssenhoff_torsion_LRS}
\end{equation}
where $\delta=\frac{1}{2}\varepsilon^{\alpha\beta}\Delta_{\alpha\beta}$,
with the constraint $\tau=\delta$.

Now, an LRS space-time is said to be of class I (LRSI) if the congruence
of the curves associated with vector field $e$ - defined to have
the same direction as the axis of symmetry - is hypersurface orthogonal.\footnote{
Following Ref.~\citep{Stewart_Ellis}, a space-time is said to be
LRS II when it has locally rotational symmetry and the vector fields
$u$ and $e$ are hypersurface orthogonal. Just so happens, in space-times
with null-torsion, an hypersurface orthogonal congruence has null
vorticity. As such, in literature, LRS II space-times are characterized
and usually referred as space-times with locally rotational symmetry
and vorticity free $u$ and $e$ vector fields. As was shown in Ref.~\citep{Luz_Mena}
this is not the case for space-times with non-null torsion where an
hypersurface orthogonal congruence does not have null vorticity. In
this article, we will follow the naming convention of Ref.~\citep{Stewart_Ellis}.
This has at least one advantage: when comparing results with the null
torsion case, we simply have to compare with the same named class;
for instance, static spherically symmetric space-times, with or without
torsion, always fall in the category of static LRS II space-times
.}
If the congruence of curves associated with the vector field $u$
is also hypersurface orthogonal, the space-time is said to be LRS
of class II (LRSII). From the results in Ref.~\citep{Luz_Mena},
for a torsion tensor given by Eq.~\eqref{eq:Weyssenhoff_torsion}
we have that $e$ will be hypersurface orthogonal if and only if 
\begin{equation}
\xi=0\,,\label{eq:hypersurface_orthogonal_consequences_e}
\end{equation}
and that $u$ will be hypersurface orthogonal if and only if 
\begin{equation}
\begin{aligned}\Omega & =\tau\,,\\
s_{\alpha} & =0\,,
\end{aligned}
\label{eq:hypersurface_orthogonal_consequences_u}
\end{equation}
where we opted to highlight that $s_{\alpha}$ will also be null from
the imposition that the congruence of $u$ is hypersurface orthogonal.

Before proceeding we should point out the fact that $s_{\alpha}=0$
has an interesting effect on the nature of the Weyssenhoff fluid.
Comparing Eq.~\eqref{eq:Weyssenhoff_energy_momentum_tensor} with
Eq.~\eqref{eq:Energy_momentum_tensor_decomposition_general} we conclude
that for a Weyssenhoff fluid the only non-null covariantly defined
quantities in Eq.~\eqref{eq:Energy_momentum_tensor_decomposition_quantities}
are $\mu$, $p$ and $q_{1\alpha}=-\left(\mathcal{A}e^{\mu}+\mathcal{A}^{\mu}\right)S_{\mu\alpha}$.
Now, since in an LRS space-time both $\mathcal{A_{\alpha}}$ and $s_{\alpha}$
are null, it implies that $q_{1\alpha}=-\left(\mathcal{A}e^{\mu}+\mathcal{A}^{\mu}\right)S_{\mu\alpha}=0$,
that is, the contributions of spin in the Weyssenhoff fluid model
for an LRS space-time, will not appear in the canonical energy-momentum
tensor. From this result, one might (wrongly) conclude that 
torsion has no role in the dynamics of the setup. In reality, torsion
will still markedly influence the behavior of the matter fields. Indeed,
for instance, when comparing to space-times with null torsion, where
LRS II space-times are necessarily irrotational (cf. e.g. Ref.~\citep{Stewart_Ellis}),
the presence of a non-null torsion of the form of Eq.~\eqref{eq:Weyssenhoff_torsion}
will induce a non-null vorticity of the congruence of curves associated
with $u$, Eq.~\eqref{eq:hypersurface_orthogonal_consequences_u}.
Thus, although in the considered setup spin does not appear in the
canonical energy-momentum tensor, it will still markedly change the
geometry of the space-time.

An additional assumption we will consider is that the space-time is
static. Now, a space-time is said to be stationary if it admits the
existence of a time-like Killing vector field $\Psi$. If the congruence
of time-like curves associated with $\Psi$ are also hypersurface
orthogonal the space-time is said to be static. Given that the choice
of the vector field $u$ is arbitrary, we can write, at each point,
$\Psi=C\,u$, where $C=C\left(x^{\alpha}\right)$ is a generic non-null
smooth function of the coordinates. The Killing equation $\mathcal{L}_{\Psi}g_{\alpha\beta}=0$
in presence of torsion can be written as 
\begin{equation}
\nabla_{\left(\alpha\right.}\Psi_{\left.\beta\right)}+2S_{\sigma\left(\alpha\beta\right)}\Psi^{\sigma}=0\,,\label{eq:Killing_equation}
\end{equation}
for any metric compatible connection. Assuming Eq.~\eqref{eq:Weyssenhoff_torsion},
contracting Eq.~\eqref{eq:Killing_equation} with $h_{\mu}^{\alpha}h_{\nu}^{\beta}$
and $h^{\alpha\beta}$ we have 
\begin{equation}
\left\{ \theta,\,\Sigma,\,\Sigma_{\alpha},\,\Sigma_{\alpha\beta}\right\} =0\,,
\end{equation}
and $u^{a}\partial_{a}C\left(x^{\alpha}\right)=0$. All is left now
is to impose the condition that $\Psi$ is hypersurface orthogonal.
However, if $u$ is hypersurface orthogonal, so is any $\Psi=C\,u$.
Hence, for the space-time to be static, Eqs. \eqref{eq:hypersurface_orthogonal_consequences_u}
must hold.

Lastly, computing the quantities: $N_{\mu}^{\alpha}N_{\nu}^{\gamma}v^{\beta}R_{\alpha\beta\gamma\delta}v^{\delta}$,
$\varepsilon_{\mu}{}^{\alpha\beta}R_{\alpha\beta\gamma\delta}v^{\delta}$
and $\varepsilon^{\mu\gamma}v^{\beta}R_{\alpha\beta\gamma\delta}e^{\delta}$,
we also find that in the considered setup 
\begin{equation}
\left\{ \mathbb{E},\,\mathcal{E}_{\alpha\beta},\,\mathcal{H}_{\alpha\beta},\bar{\mathcal{H}}_{\alpha\beta}\right\} =0\,.
\end{equation}
Therefore, gathering the previous results we find that stationary
LRS I or LRS II space-times permeated by an uncharged Weyssenhoff
fluid are characterized by the following set of quantities $\left\{ \mu,\,p,\,\phi,\,\Omega,\,\mathcal{A},\,\tau,\,\mathcal{E},\,\mathrm{H},\,\bar{\mathrm{H}},\,\mathcal{H},\,\bar{\mathcal{H}}\right\} $.

\subsection{Structure equations}

We are now in position to find the structure equations for stationary,
locally rotationally symmetric space-time filled by a Weyssenhoff
fluid in the case where the congruence of space-like curves associated
with $e$ are hypersurface orthogonal, that is, in the case when $\xi=0$.
The non-trivial, independent propagation equations are 
\begin{align}
\hat{p}+\mathcal{A}\left(\mu+p\right) & =-\frac{1}{4\pi}\tau\bar{\mathcal{H}}\,,\label{eq:SE_p_hat}\\
\hat{\mathcal{A}}+\mathcal{A}\left(\mathcal{A}+\phi\right)+2\Omega^{2} & =4\pi\left(\mu+3p\right)\,,\label{eq:SE_A_hat}\\
\hat{\phi}+\frac{1}{2}\phi^{2}+\mathcal{E} & =-\frac{16\pi}{3}\mu\,,\label{eq:SE_phi_hat}\\
\hat{\mathcal{E}}+\frac{3}{2}\mathcal{E}\phi+\Omega\mathrm{H}+2\left(\tau-\Omega\right)\bar{\mathcal{H}} & =\frac{8\pi}{3}\hat{\mu}\,,\label{eq:SE_E_hat}\\
\hat{\mathcal{H}}-\frac{1}{2}\phi\left(\mathrm{H}-2\mathcal{H}\right)+\mathcal{E}\left(3\Omega-2\tau\right) & =-8\pi\Omega\left(\mu+p\right)+\nonumber \\
 & +\frac{8\pi}{3}\tau\left(\mu+3p\right)\,,\label{eq:SE_H1_hat}\\
2\hat{\Omega}+\Omega\phi & =\mathrm{H}\,,\label{eq:SE_constraint_H_1}
\end{align}
and the constraint equations
\begin{align}
\mathcal{E}+\mathcal{A}\phi+2\Omega^{2} & =\frac{8\pi}{3}\left(\mu+3p\right)\,,\label{eq:SE_constraint_E_A_Omega}\\
2\mathcal{A}\left(\Omega-\tau\right)-\Omega\phi+\mathcal{H} & =0\,,\label{eq:SE_constraint_curly_H_1}\\
\Omega\left(\phi-2\mathcal{A}\right)+\bar{\mathrm{H}} & =0\,,\label{eq:SE_constraint_H_2}\\
\mathrm{H}+\bar{\mathrm{H}}+2\bar{\mathcal{H}} & =0\,.\label{eq:SE_constraint_curly_H_2}
\end{align}

Let us discuss the cases when the congruence associated with $u$
is either hypersurface orthogonal or not, separately. Consider the
cases when $\Omega\neq\tau$. In such cases we find from Eqs.~\eqref{eq:SE_p_hat}
- \eqref{eq:SE_constraint_curly_H_2} the following relation between
$\Omega$ and $\tau$ 
\begin{equation}
\left(\phi-\mathcal{A}\right)\left(\tau-\Omega\right)+\hat{\tau}-\hat{\Omega}=0\,,\label{eq:SE_Omega_hat_tau_hat}
\end{equation}
leading us to conclude that the difference between $\tau$ and $\Omega$
can be uniquely described by the behavior of the variables $\phi$
and $\mathcal{A}$. Notice that if at an initial instant $\Omega$
and $\tau$ are different then, unless the term $\phi-A$ diverges,
there will be no point in which they are equal. Conversely, if $\Omega=\tau$
at a point these two quantities will be equal at any point.

The relations $\Omega=\tau$ or Eq.~\eqref{eq:SE_Omega_hat_tau_hat},
for stationary LRS II or LRS I space-times, respectively, have the
advantage of not depending directly of the magnetic components of
the Weyl tensor and they can replace one of Eqs.~\eqref{eq:SE_p_hat}
- \eqref{eq:SE_constraint_curly_H_2}. As we shall see, it is useful
to remove Eq.~\eqref{eq:SE_H1_hat}.

Finally, to close the system we will need an equation of state that
relates the pressure of the fluid with its energy-density: $p=p\left(\mu\right)$;
and an equation that relates the energy-density of the fluid with
the intrinsic hypermomentum: $\delta=\delta\left(p(\mu),\mu\right)$.

\section{Generalized TOV equation for stationary LRS I and LRS II space-times\label{sec:Generalized_TOV_equation}}

With the full set of structure equations we are finally in position
to make the derivation of the generalized Tolman-Oppenheimer-Volkoff
(TOV) equations. Let us start by introducing the scalar function 
\begin{equation}
K=\frac{8\pi}{3}\mu-\mathcal{E}+\frac{1}{4}\phi^{2}-3\Omega^{2}+2\Omega\tau\,,\label{eq:Definition_K}
\end{equation}
with the following property 
\begin{equation}
\hat{K}=-\phi K\,,\label{eq:Propagation_K}
\end{equation}
found from the structure equations. Eq.~\eqref{eq:Definition_K}
generalizes the expressions in Refs.~\citep{Betschart,Burston} \footnote{Notice that in Ref.~\citep{Burston} there is a small typographic
error.}. Moreover, since the Gauss equation is unchanged by the presence
of torsion, it is possible to prove that, in the cases when the vector
fields $u$ and $e$ are hypersurface orthogonal, the quantity $K$
represents the Gaussian curvature of the 2-sheet orthogonal to both
$u$ and $e$.

Now, following the treatment in Refs.~\citep{Sante1,Sante2}, without
loss of generality, let us re-parameterize the integral curves of
$e$ using a, in general non-affine, parameter $\rho$, such that
for an arbitrary scalar function $F$ 
\begin{equation}
\hat{F}=\phi F_{,\rho}\,.\label{eq:Definition_parameter_rho}
\end{equation}
In particular we have 
\begin{equation}
K_{,\rho}=-K\,.\label{eq:Propagation_K_rho}
\end{equation}

Introducing the following set of variables 
\begin{equation}
\begin{aligned}\mathbb{X} & =\frac{\phi_{,\rho}}{\phi}\,, & \; &  & \mathbb{B}_{1} & =\frac{\mathrm{H}}{\phi^{2}}\,, & \; &  & \mathcal{M} & =8\pi\frac{\mu}{\phi^{2}}\,,\\
\mathbb{Y} & =\frac{\mathcal{A}}{\phi}\,, &  &  & \mathbb{B}_{2} & =\frac{\bar{\mathrm{H}}}{\phi^{2}}\,, &  &  & \mathcal{P} & =8\pi\frac{p}{\phi^{2}}\,,\\
\mathbb{E} & =\frac{\mathcal{E}}{\phi^{2}}\,, &  &  & \mathbb{D}_{1} & =\frac{\mathcal{H}}{\phi^{2}}\,, &  &  & \Delta & =\frac{\delta}{\phi}\,,\\
\mathbb{T} & =\frac{\tau}{\phi}\,, &  &  & \mathbb{D}_{2} & =\frac{\bar{\mathcal{H}}}{\phi^{2}}\,,\\
\mathcal{\mathbb{W}} & =\frac{\Omega}{\phi}\,, &  &  & \mathcal{K} & =\frac{K}{\phi^{2}}\,,
\end{aligned}
\label{eq:BlackBoard_quantities_definition}
\end{equation}
we can re-write Eqs.~\eqref{eq:SE_p_hat} - \eqref{eq:SE_constraint_curly_H_2}
as 
\begin{align}
2\mathbb{Y}_{,\rho}+2\mathbb{Y}\left(\mathbb{X}+\mathbb{Y}+1\right) & =\mathcal{M}+3\mathcal{P}-4\mathbb{W}^{2}\,,\label{eq:Alt_SE_prop_Y}\\
\mathcal{K}_{,\rho}+\mathcal{K}\left(2\mathbb{X}+1\right) & =0\,,\label{eq:Alt_SE_prop_K}\\
\mathcal{P}_{,\rho}+\mathcal{P}\left(2\mathbb{X}+\mathbb{Y}\right)+\mathbb{Y}\mathcal{M} & =2\mathbb{T}\mathbb{W}\left(\mathbb{X}+\mathbb{Y}\right)+2\mathbb{T}\mathbb{W}_{,\rho}\,,\label{eq:Alt_SE_prop_P}\\
2\mathbb{W}_{,\rho}+\mathbb{W}\left(2\mathbb{X}+1\right) & =\mathbb{B}_{1}\,,\label{eq:Alt_SE_prop_O}
\end{align}
with the constraints
\begin{align}
\mathcal{M}+3\mathcal{P}-3\mathbb{Y}-3\mathbb{E}-6\mathbb{W}^{2} & =0\,,\label{eq:Alt_SE_constraint_MPYEO}\\
2\mathcal{M}+2\mathbb{X}+2\mathcal{P}-2\mathbb{Y}-4\mathbb{W}^{2}+1 & =0\,,\label{eq:Alt_SE_constraint_MXPYO}\\
4\mathbb{Y}+4\mathbb{W}\left(2\mathbb{T}-\mathbb{W}\right)-4\mathcal{P}-4\mathcal{K}+1 & =0\,,\label{eq:Alt_SE_constraint_YOTPK}\\
\mathbb{D}_{1}+\mathcal{\mathbb{W}}\left(2\mathbb{Y}-1\right)-2\mathbb{Y}\mathbb{T} & =0\,,\label{eq:Alt_SE_constraint_DOYT}\\
\mathbb{B}_{2}+\mathbb{W}\left(1-2\mathbb{Y}\right) & =0\,,\\
\mathbb{B}_{1}+\mathbb{B}_{2}+2\mathbb{D}_{2} & =0\,,\label{eq:Alt_SE_constraint_B_D}\\
\mathbb{T} & =\Delta\,,\label{eq:Alt_SE_constraint_t_delta}
\end{align}
and, depending on whether we are considering stationary LRS I or LRS
II space-times, we have the extra equation 
\begin{equation}
\begin{cases}
\mathbb{W}_{,\rho}-\mathbb{T}_{,\rho}=\left(1-\mathbb{Y}+\mathbb{X}\right)\left(\mathbb{T}-\mathbb{W}\right) & \text{, if LRS I},\\
\mathbb{W}=\mathbb{T} & \text{, if LRS II}.
\end{cases}\label{eq:Alt_SE_T_O_relation}
\end{equation}
The system is closed provided and equation of state such that $\mathcal{P}=\mathcal{P}\left(\mathcal{M}\right)$
and a relation such that $\Delta=\Delta\left(\mathcal{P}(\mathcal{M}),\mathcal{M}\right)
$.

Now, using Eqs.~\eqref{eq:Alt_SE_constraint_MXPYO} and \eqref{eq:Alt_SE_constraint_YOTPK}
to eliminate $\mathbb{X}$ and $\mathbb{Y}$ in Eqs.~\eqref{eq:Alt_SE_prop_P}
and \eqref{eq:Alt_SE_T_O_relation} we find
\begin{equation}
\begin{aligned}\mathcal{P}_{,\rho} & =-\mathcal{P}^{2}+\mathcal{P}\left[\frac{7}{4}-3\mathcal{K}+\mathcal{\mathbb{W}}\left(8\Delta-7\mathcal{\mathbb{W}}\right)\right]+\\
 & +2\Delta\left[\left(2\mathcal{K}-1\right)\mathcal{\mathbb{W}}-4\mathcal{\mathbb{W}}^{2}\left(\Delta-\mathcal{\mathbb{W}}\right)+\mathcal{\mathbb{W}}_{,\rho}\right]+\\
 & +\mathcal{M}\left(\frac{1}{4}-\mathcal{K}+\mathcal{P}-\mathcal{\mathbb{W}}^{2}\right)\,,\\
\mathcal{K}_{,\rho} & =-2\mathcal{K}\left(3\mathcal{\mathbb{W}}^{2}-2\Delta\mathcal{\mathbb{W}}+\mathcal{K}-\mathcal{M}-\frac{1}{4}\right)\,,\\
\mathcal{P} & =\mathcal{P\left(M\right)}\,,\\
\Delta & =\Delta\left(\mathcal{P}(\mathcal{M}),\mathcal{M}\right)\,,
\end{aligned}
\label{eq:TOV_general}
\end{equation}
and

\begin{equation}
\begin{cases}
\begin{aligned}\mathcal{\mathbb{W}}_{,\rho}-\Delta_{,\rho} & =\frac{1}{2}\left(\Delta-\mathcal{\mathbb{W}}\right)\times\\
 & \hspace{0.4cm}\times\left(1-2\mathcal{M}-2\mathcal{P}+4\mathcal{\mathbb{W}}^{2}\right)
\end{aligned}
 & \text{, if LRS I}\\
\mathcal{\mathbb{W}}=\Delta & \text{, if LRS II}
\end{cases}
\end{equation}
which represents the covariant TOV equations. The system is completed
by the extra relations:
\begin{align}
\mathcal{K}-\frac{1}{4}+\mathcal{P}+\mathcal{\mathbb{W}}\left(\mathcal{\mathbb{W}}-2\Delta\right) & =\mathbb{Y}\,,\\
\mathcal{K}-\frac{3}{4}-\mathcal{M}+\mathcal{\mathbb{W}}\left(3\mathcal{\mathbb{W}}-2\Delta\right) & =\mathbb{X}\,,\\
\mathcal{M}+\mathcal{\mathbb{W}}\left(6\Delta-9\mathcal{\mathbb{W}}\right)-3\left(\mathcal{K}-\frac{1}{4}\right) & =3\mathbb{E}\,,\\
2\mathcal{\mathbb{W}}_{,\rho}+\mathcal{\mathbb{W}}\left(2\mathcal{K}-2\mathcal{M}-4\Delta\mathcal{\mathbb{W}}+6\mathcal{\mathbb{W}}^{2}-\frac{1}{2}\right) & =\mathbb{B}_{1}\,,\\
\mathcal{\mathbb{W}}\left(2\mathcal{K}+2\mathcal{P}-4\Delta\mathcal{\mathbb{W}}+2\mathcal{\mathbb{W}}^{2}-\frac{3}{2}\right) & =\mathbb{B}_{2}\,,\\
\mathbb{B}_{1}+\mathbb{B}_{2}+2\mathbb{D}_{2} & =0\,,\\
\nonumber \\
\mathcal{\mathbb{W}}\left(6\Delta\mathcal{\mathbb{W}}-2\mathcal{\mathbb{W}}^{2}-4\Delta^{2}+1\right)+\hspace{0.5cm}\nonumber \\
+2\left(\Delta-\mathcal{\mathbb{W}}\right)\left(\mathcal{K}+\mathcal{P}-\frac{1}{4}\right) & =\mathbb{D}_{1}\label{eq:TOV_D1_general}
\end{align}

\subsection{The static case\label{subsec:The-static-case}}

The full set of Eqs.~\eqref{eq:TOV_general} - \eqref{eq:TOV_D1_general}
completely describe the geometry of a stationary LRS I or LRS II space-time
filled by an Weyssenhoff fluid. Let us now consider the particular
cases when the space-time is static, that is the case when $\mathbb{W=T=}\Delta$.

Introducing the following quantities 
\begin{equation}
\begin{aligned}\mathscr{M} & =\mathcal{M}-\Delta^{2}\,,\\
\mathscr{P} & =\mathcal{P}-\Delta^{2}\,,\\
\mathscr{E} & =\mathbb{E}+\frac{2}{3}\Delta^{2}\,,
\end{aligned}
\label{eq:TOV_static_effective_quantities}
\end{equation}
Eqs\@.~\eqref{eq:TOV_general} are given by 
\begin{align}
\mathscr{P}_{,\rho} & =-\mathscr{P}^{2}+\mathscr{P}\left[\mathscr{M}+1-3\left(\mathcal{K}-\frac{1}{4}\right)\right]-\nonumber \\
 & \hspace{0.4cm}-\mathscr{M}\left(\mathcal{K}-\frac{1}{4}\right)\,,\label{eq:TOV_static_P}\\
\mathcal{K}_{,\rho} & =-2\mathcal{K}\left(\mathcal{K}-\frac{1}{4}-\mathscr{M}\right)\,.\label{eq:TOV_static_K}\\
\mathbb{Y} & =\mathcal{K}-\frac{1}{4}+\mathscr{P}\,,\label{eq:TOV_static_Y}\\
\mathbb{X} & =\mathcal{K}-\frac{3}{4}-\mathscr{M}\,,\label{eq:TOV_static_X}\\
3\mathscr{E} & =\mathscr{M}-3\left(\mathcal{K}-\frac{1}{4}\right)\,,\label{eq:TOV_static_E}
\end{align}
which match exactly the expressions found in the theory of General
Relativity ( cf. Ref.~\citep{Sante1}) for an effective energy density
and pressure and the corrected electric part of the Weyl tensor: $\mathscr{M}$,$\mathscr{P}$
and $\mathscr{E}$. Note that the extra constraints for the magnetic
components of the Weyl tensor 
\begin{align}
2\Delta_{,\rho}+2\Delta\left(\mathcal{K}-\mathscr{M}-\frac{1}{4}\right) & =\mathbb{B}_{1}\,,\label{eq:TOV_static_B1}\\
\mathbb{B}_{2}+\Delta\left[1-2\mathscr{P}-2\left(\mathcal{K}-\frac{1}{4}\right)\right] & =0\,,\\
\mathbb{B}_{1}+\mathbb{B}_{2}+2\mathbb{D}_{2} & =0\,,\\
\mathbb{D}_{1} & =\Delta\,,\label{eq:TOV_static_D1}
\end{align}
imply that the geometry of the space-time is fundamentally different
from the corresponding one in General Relativity. Nonetheless, the
fact that Eqs.~\eqref{eq:TOV_static_P} - \eqref{eq:TOV_static_E}
have the same form for the corrected quantities in Eq.~\eqref{eq:TOV_static_effective_quantities}
lead us to the notable result: 
\begin{prop}
At the level of the metric, all static, locally rotationally symmetric
of class II solutions of the theory of General Relativity for a perfect
fluid with energy momentum $\left(\mathcal{T}^{GR}\right)_{\alpha\beta}=\mu\,u_{\alpha}u_{\beta}+ph_{\alpha\beta}$,
are also solutions of Einstein-Cartan theory sourced by a Weyssenhoff
fluid with energy-momentum tensor $\left(\mathcal{T}^{EC}\right)_{\alpha\beta}=\left(\mu+\frac{\delta^{2}}{8\pi}\right)\,u_{\alpha}u_{\beta}+\left(p+\frac{\delta^{2}}{8\pi}\right)h_{\alpha\beta}$. 
\end{prop}

It is important to stress that, because of the nature of the corrections
in Eq.~\eqref{eq:TOV_static_effective_quantities}, solutions which
are unacceptable in General Relativity due, for example, to negative
energy densities or pressure, might still correspond to physically
acceptable ones in the Einstein-Cartan case.

In the rest of the article we will consider the case of static spherically
symmetric space-times, hence, we will study solutions of Eqs.~\eqref{eq:TOV_static_effective_quantities}
- \eqref{eq:TOV_static_D1}.

\section{Junction Conditions\label{sec:Junction_Conditions}}

In the analysis of compact objects in a geometric theory of gravity
it is often necessary to model the space-time as two distinct manifolds
glued together at a common boundary. Such operation is usually performed
using Israel junction conditions \citep{Israel}. The Israel procedure
was initially developed for space-times in the absence of torsion.
More recently, a number of works have been published in which the
junction conditions are generalized to space-times with torsion in
different contexts \citep{AKP,Vignolo}. Here we will summarize the
results and extend them in light of the structure equations we have
just obtained.

Consider two Lorentzian manifolds with boundary: $\mathcal{V}^{-}$
and $\mathcal{V}^{+}$, matched at an hypersurface $\mathcal{N}$,
forming a new manifold $\mathcal{V}$. Let $n$ represent the unit
normal to $\mathcal{N}$, pointing from $\mathcal{V}^{-}$ to $\mathcal{V}^{+}$,
and $e_{a}$ be the tangent vectors to $\mathcal{N}$. Here $\mathcal{N}$
can be either time-like or space-like. Now, following Ref.~\citep{Vignolo},
for the total space-time to be a valid solution of the the field equations
and to guarantee that at $\mathcal{N}$ there is no surface layer,
the following conditions must be met: 
\begin{itemize}
\item the induced metric at $\mathcal{N}$, as seen from each space-time
$\mathcal{V}^{-}$ and $\mathcal{V}^{+}$, $h_{ab}^{\pm}:=g_{\alpha\beta}^{\pm}e_{a}^{\alpha}e_{b}^{\beta}$,
must be the same, 
\begin{equation}
\left[h_{ab}\right]_{\pm}=0\,;\label{eq:Junction_cond_1}
\end{equation}
\item the extrinsic curvature tensor of $\mathcal{N}$ as seen from $\mathcal{V}^{-}$
and $\mathcal{V}^{+}$, $Q_{ab}^{\pm}:=e_{a}^{\alpha}e_{b}^{\beta}\nabla_{\alpha}^{\pm}n_{\beta}$
, is such that 
\begin{equation}
\left[Q_{ab}\right]_{\pm}=0\,;\label{eq:Junction_cond_2}
\end{equation}
\item the torsion tensor verifies 
\begin{equation}
h^{\alpha\beta}\left[S_{\mu\beta}{}^{\mu}\right]_{\pm}+\epsilon h^{\alpha\beta}\left[n^{\mu}n^{\nu}S_{\beta\mu\nu}\right]_{\pm}=0\,.\label{eq:Junction_cond_3}
\end{equation}
\end{itemize}
For simplicity we labeled a field $\Upsilon$ defined on the the sub-manifold
$\mathcal{V}^{+}$ or $\mathcal{V}^{-}$ as $\Upsilon_{+}\equiv\Upsilon\left(\mathcal{V}^{+}\right)$
or $\Upsilon_{-}\equiv\Upsilon\left(\mathcal{V}^{-}\right)$, respectively
and use the notation $\left[\Upsilon\right]_{\pm}$ to represent the
difference of a field as measured from each sub-manifold at the matching
surface, i.e., $\left[\Upsilon\right]_{\pm}\equiv\left.\Upsilon\left(\mathcal{V}^{+}\right)\right|_{\mathcal{N}}-\left.\Upsilon\left(\mathcal{V}^{-}\right)\right|_{\mathcal{N}}$.

Clearly, conditions \eqref{eq:Junction_cond_1} - \eqref{eq:Junction_cond_3}
reduce to the Israel conditions \citep{Israel} in the limit of null
torsion. In that case, the junction conditions not only guarantee
that at the matching surface there is no thin-shell but also are necessary
and sufficient to guarantee that the singular part of the Riemann
tensor is null. However, this is not the case for torsional space-times.
In the presence of torsion, assuming only compatibility with the metric,
the Riemann tensor of the total space-time is given by\footnote{Here we mix the distribution associated to a tensor and the tensor
itself which is, strictly speaking, an abuse of language. Our conclusions,
however, are not influence by this issue. See e.g. Ref.~\citep{Senovilla}
and references therein for more details.} 
\begin{align}
R_{\alpha\beta\gamma}{}^{\rho} & =\theta\left(\lambda\right)R_{\alpha\beta\gamma}^{+}{}^{\rho}+\theta\left(-\lambda\right)R_{\alpha\beta\gamma}^{-}{}^{\rho}+\nonumber \\
 & \hspace{0.4cm}+\delta\left(\lambda\right)\left(A_{\alpha\beta\gamma}{}^{\rho}+B_{\alpha\beta\gamma}{}^{\rho}\right)\,,
\end{align}
where $\theta\left(\lambda\right)$ represents the Heaviside distribution,
$\delta\left(\lambda\right)$ the Dirac distribution, $\lambda$ is
the parameter of the integral curves of $n$, adjusted such that the
matching surface is located at $\lambda=0$, $R_{\alpha\beta\gamma}^{\pm}{}^{\rho}$
represent the Riemann tensors of the $\mathcal{V}^{-}$ and $\mathcal{V}^{+}$
sub-manifolds and 
\begin{align}
A_{\alpha\beta\gamma}{}^{\rho} & =\epsilon\left(n_{\beta}\left[\Gamma_{\alpha\gamma}^{\rho}\right]_{\pm}-n_{\alpha}\left[\Gamma_{\beta\gamma}^{\rho}\right]_{\pm}\right)\,,\label{eq:Junction_Riemann_sing_A}\\
B_{\alpha\beta\gamma}{}^{\rho} & =\epsilon\left(n_{\beta}\left[K_{\alpha\gamma}{}^{\rho}\right]_{\pm}-n_{\alpha}\left[K_{\beta\gamma}{}^{\rho}\right]_{\pm}\right)\,,\label{eq:Junction_Riemann_sing_B}
\end{align}
are the singular parts of the - total - Riemann tensor, with $\Gamma_{\alpha\gamma}^{\rho}$
being the Christoffel symbols, $K_{\alpha\beta}{}^{\gamma}\equiv S_{\alpha\beta}{}^{\gamma}+S^{\gamma}{}_{\alpha\beta}-S_{\beta}{}^{\gamma}{}_{\alpha}$
the contorsion tensor and $\epsilon=n_{\mu}n^{\mu}$. We see that
in general the conditions \eqref{eq:Junction_cond_1} - \eqref{eq:Junction_cond_3}
do not guarantee that both the tensors in Eq.~\eqref{eq:Junction_Riemann_sing_A}
and \eqref{eq:Junction_Riemann_sing_B} are null. A smooth junction
of two space-times has to imply that the discontinuities of all curvature
tensors across the matching surface have to be at most finite otherwise,
the space-time will be singular at $\mathcal{N}$. In the torsion
free case, no condition on the tensor \eqref{eq:Junction_Riemann_sing_B}
is required as it is identically zero and therefore does not appear
in the Einstein equations. In the Eistein-Cartan case, however, even
imposing Eqs.~\eqref{eq:Junction_cond_1} - \eqref{eq:Junction_cond_3},
the remaining singular part of the Riemann tensor will appear in the
structure equations, leading to a singularity in $\mathcal{N}$. Differently
from the standard violation of Israel's condition, such singularity
can not be attributed to the presence of a thin shell since, Eqs.~\eqref{eq:Junction_cond_1}
- \eqref{eq:Junction_cond_3} prevent the existence of a surface layer
at $\mathcal{N}$. For this reason, in the following we will require
a completely smooth matching of the Riemann tensor on the boundary. It is a known result (see e.g. \citep{Senovilla,Poisson} for a clear
derivation) that the tensor $A_{\alpha\beta\gamma}{}^{\rho}$, Eq.~\eqref{eq:Junction_Riemann_sing_A},
is null if and only if $\left[K_{ab}\right]_{\pm}=0$, where $K_{ab}:=e_{a}^{\alpha}e_{b}^{\beta}\tilde{\nabla}_{\alpha}n_{\beta}$
represents the extrinsic curvature computed from the metric connection.
On the other hand, a necessary and sufficient condition for $B_{\alpha\beta\gamma}{}^{\rho}$
to be null is given by 
\begin{equation}
\left[K_{\alpha\beta}{}^{\rho}\right]_{\pm}=\epsilon n_{\alpha}\left[n^{\mu}K_{\mu\beta}\,^{\rho}\right]_{\pm}.
\end{equation}

Thus we the arrive to the following proposition: 
\begin{prop}
Let $\mathcal{V}^{-}$ and $\mathcal{V}^{+}$ be two Lorentzian manifolds
with boundary, endowed with a metric compatible, affine connection.
$\mathcal{V}^{-}$ and $\mathcal{V}^{+}$ can be smoothly matched
at a common, non-null, hypersurface $\mathcal{N}$ when the following
three conditions are verified:
\begin{enumerate}[itemsep=10pt,labelindent=\parindent,leftmargin=*,label=\roman*.,widest=II,align=left ]
\item the induced metric at $\mathcal{N}$ is such that 
\begin{equation}
\left[h_{ab}\right]_{\pm}=0\,;\label{eq:Junction_smooth_cond_1}
\end{equation}
\item the jump of extrinsic curvature of $\mathcal{N}$ is null, that is
\begin{equation}
\left[Q_{ab}\right]_{\pm}=0\,;\label{eq:Junction_smooth_cond_2}
\end{equation}
\item the jump of the contorsion tensor at $\mathcal{N}$ verifies 
\begin{equation}
\left[K_{\alpha\beta}{}^{\rho}\right]_{\pm}=\epsilon n_{\alpha}\left[n^{\mu}K_{\mu\beta}\,^{\rho}\right]_{\pm}\,,\label{eq:Junction_smooth_cond_3}
\end{equation}
where the vector field $n$ represents the unit normal to $\mathcal{N}$. 
\end{enumerate}
\end{prop}

Notice that writing $Q_{ab}=K_{ab}-e_{a}^{\alpha}e_{b}^{\beta}K_{\alpha\beta}{}^{\gamma}n_{\gamma}$
and using Eq.~\eqref{eq:Junction_smooth_cond_3}, Eq.~\eqref{eq:Junction_smooth_cond_2}
is the same as imposing $\left[K_{ab}\right]_{\pm}=0$.

\subsection{1+1+2 junction of static LRSII space-times with torsion\label{1+1+2Junction}}

Let us now express conditions \eqref{eq:Junction_smooth_cond_1} -
\eqref{eq:Junction_smooth_cond_3} covariantly in the specific case
of two static LRS II space-times endowed with a torsion tensor field
of the form $S_{\alpha\beta}\,^{\gamma}=\varepsilon_{\alpha\beta}u^{\gamma}\tau$,
where $\tau$ is a generic function of the space-time coordinates.

In what follows we will be interested in the case when the interior
and exterior space-times are to be matched at a time-like hypersurface,
orthogonal to the vector field $e$. Then, condition \eqref{eq:Junction_smooth_cond_1}
reads 
\begin{equation}
\left[N_{\alpha\beta}-u_{\alpha}u_{\beta}\right]_{\pm}=0\,,\label{eq:Junction_cond_particular_1}
\end{equation}
where $N_{\alpha\beta}$ verifies Eqs.~\eqref{eq:projector_N_properties}.
Using Eq.~\eqref{eq:Cov_vector_e}, in the considered setup, Eq.~\eqref{eq:Junction_smooth_cond_2}
is simply 
\begin{equation}
\left[\frac{1}{2}\phi N_{\alpha\beta}-\mathcal{A}u_{\alpha}u_{\beta}\right]_{\pm}=0\,,
\end{equation}
which, contracting with the induced metric at $\mathcal{N}$ and using
Eq.~\eqref{eq:Junction_cond_particular_1}, gives 
\begin{equation}
\left[\phi-\mathcal{A}\right]_{\pm}=0\,.\label{eq:Junction_cond_particular_3}
\end{equation}
From Eqs.~\eqref{eq:Junction_cond_particular_1} - \eqref{eq:Junction_cond_particular_3}
we find that at the matching surface the following constraints have
to be met 
\begin{align}
\left[\phi\right]_{\pm} & =0\,,\label{eq:Junction_cond_particular_phi}\\
\left[\mathcal{A}\right]_{\pm} & =0\,,\label{eq:Junction_cond_particular_A}
\end{align}
implying, for $\phi\neq0$, 
\begin{equation}
\left[\mathbb{Y}\right]_{\pm}=0\,.
\end{equation}
Given that $e$ is continuous across $\mathcal{N}$, we can integrate
Eq.~\eqref{eq:Propagation_K_rho}, finding $K=k_{0}e^{-\rho}$. Using
Eq.~\eqref{eq:BlackBoard_quantities_definition} and \eqref{eq:Junction_cond_particular_phi}
we have 
\begin{equation}
\left[\mathcal{K}\right]_{\pm}=0\,.\label{eq:Junction_cond_particular_K}
\end{equation}
Using the previous results in Eq.~\eqref{eq:TOV_static_Y} we arrive
at 
\begin{equation}
\left[8\pi p-\delta^{2}\right]_{\pm}=0\,.\label{eq:Junction_cond_particular_P_effective}
\end{equation}

Finally, for the specific type of torsion that we consider in this
article, condition \eqref{eq:Junction_smooth_cond_3} imposes 
\begin{equation}
\left[\delta\right]_{\pm}=0\,,\label{eq:Junction_cond_particular_spin_density}
\end{equation}
then, from Eq.~\eqref{eq:Junction_cond_particular_P_effective},
\begin{equation}
\left[p\right]_{\pm}=0\,.\label{eq:Junction_cond_particular_pressure}
\end{equation}

We have then found that for a smooth matching between two static LRS
II space-times endowed with a torsion tensor field of the form $S_{\alpha\beta}\,^{\gamma}=\varepsilon_{\alpha\beta}u^{\gamma}\tau$,
both the pressure of the fluid and the spin density, as seen from
each space-time, must match at $\mathcal{N}$.

\section{Exact solutions for static LRS II space-times\label{sec:Exact_solutions}}

Given the set of structure equations \eqref{eq:TOV_static_effective_quantities}
- \eqref{eq:TOV_static_D1} that describe the behavior of a static,
LRSII space-time filled by a Weyssenhoff fluid, let us now find and
study some exact solutions.

As was stated before, the system of structure equations is not closed
until an equation of state and an expression for the spin density
are provided. Let us then consider some particular relations for the
pressure, energy and spin densities of the fluid in order to gain
some insight into the behavior of compact objects in a fully relativistic
theory with non-null spin.

For the remaining of the article we will consider only the particular
case of spherically symmetric space-times. Moreover, in what follows
we will refer to static, spherically symmetric compact objects as
``stars''. Although this is an abuse of language, it is also a trend in the
the literature since such systems are expected to be a good model
for slowly varying astrophysical bodies.

\subsection{Effective constant energy-density and the Buchdahl limit}

We start by considering the case of a system where the effective energy
density is assumed to be constant, that is 
\begin{equation}
8\pi\mu-\delta^{2}=\tilde{\mu}_{0}\,,\label{eq:ESolutions_constant_energy_assumption}
\end{equation}
where $\tilde{\mu}_{0}\in\mathbb{R}$. Notice that, contrary to the
case of null torsion, the above assumption does not have to imply
that the energy density, $\mu$, is constant.

Using Eqs.~\eqref{eq:Propagation_K} and \eqref{eq:Definition_parameter_rho}
we have 
\begin{equation}
K\left(\rho\right)=\frac{e^{-\rho}}{r_{0}^{2}}\,,\label{eq:ESolutions_constant_energy_solutions_k}
\end{equation}
where $r_{0}$ is an integration constant. Eq.~\eqref{eq:ESolutions_constant_energy_solutions_k}
then yields 
\begin{equation}
\mathscr{M}\left(\rho\right)=\tilde{\mu}_{0}r_{0}^{2}e^{\rho}\mathcal{K}\left(\rho\right)\,.\label{eq:ESolutions_constant_energy_solutions_M}
\end{equation}
Eq.~\eqref{eq:ESolutions_constant_energy_solutions_M} allow us to
solve Eq.~\eqref{eq:TOV_static_K}, finding 
\begin{equation}
\mathcal{K}\left(\rho\right)=\frac{3}{12-4\tilde{\mu}_{0}r_{0}^{2}\,e^{\rho}+3\mathcal{K}_{0}e^{-\frac{\rho}{2}}}\,,
\end{equation}
where $\mathcal{K}_{0}$ is yet another integration constant. Setting
$\mathcal{K}_{0}=0$ to avoid a conical singularity at $\rho\to-\infty$
\citep{Wald}, the structure equations yield 
\begin{align}
p\left(\rho\right)-\frac{\delta\left(\rho\right)^{2}}{8\pi} & =-\frac{\tilde{\mu}_{0}\left(P_{0}+3\sqrt{3-\tilde{\mu}_{0}r_{0}^{2}e^{\rho}}\right)}{24\pi\left(P_{0}+\sqrt{3-\tilde{\mu}_{0}r_{0}^{2}e^{\rho}}\right)}\,,\label{eq:ESolution_constant_energy_solutions_p}\\
\mathcal{A}\left(\rho\right) & =-\frac{\tilde{\mu}_{0}r_{0}e^{\frac{\rho}{2}}}{\sqrt{3}\left(P_{0}+\sqrt{3-\tilde{\mu}_{0}r_{0}^{2}e^{\rho}}\right)}\,,\label{eq:ESolution_constant_energy_solutions_A}\\
\phi\left(\rho\right) & =\frac{2}{r_{0}\sqrt{3}}e^{-\frac{\rho}{2}}\sqrt{3-\tilde{\mu}_{0}r_{0}^{2}e^{\rho}}\,,\label{eq:ESolution_constant_energy_solutions_phi}\\
\mathcal{E} & =-\frac{2}{3}\delta^{2}\,,\label{eq:ESolution_constant_energy_solutions_E}
\end{align}
where we have chosen the direction of $e$ so that $\phi$ is positive,
and the value of the integration constant $P_{0}$ is to be determined
by the boundary conditions.

Let us now assume that relations \eqref{eq:ESolutions_constant_energy_assumption}
- \eqref{eq:ESolution_constant_energy_solutions_E} describe the interior
of a compact object matched at a boundary $\mathcal{N}$ to an exterior
space-time modeled by the Schwarzschild vacuum solution. From Eqs.~\eqref{eq:Junction_cond_particular_spin_density}
and \eqref{eq:Junction_cond_particular_pressure} we find that the
quantity in Eq.~\eqref{eq:ESolution_constant_energy_solutions_p}
must be zero at the boundary. Setting, without loss of generality,
the matching hypersurface to be at $\rho=0$, we find 
\begin{equation}
P_{0}=-3\sqrt{3-\tilde{\mu}_{0}r_{0}^{2}}\,.\label{eq:ESolution_constant_energy_solutions_boundary_P0}
\end{equation}
The matching conditions, Eqs.~\eqref{eq:Junction_cond_particular_phi}
and \eqref{eq:Junction_cond_particular_K}, imply that interior and
exterior observers agree on the value of circumferential radius of
$\mathcal{N}$, say $r_{0}$, and the Schwarzschild parameter, $M$,
is given by 
\begin{equation}
M=\frac{\tilde{\mu}_{0}r_{0}^{3}}{6}\,.\label{eq:ESolution_constant_energy_Sch_mass}
\end{equation}
Moreover, from condition \eqref{eq:Junction_cond_particular_spin_density}
we find that the spin density must go to zero at the matching surface,
that is, $\delta\left(\rho=0\right)=0$.

Given the previous results we are now in position to study some effects
arising from the presence of spin in compact objects. In the remaining
of this subsection, for clarity, we shall write the results in terms
of the circumferential radius $r$. Using the fact that, in the considered
setup, the quantity $K$, Eq.~\eqref{eq:Definition_K}, represents
the Gaussian curvature of the 2-sheet, we have that the parameter
$\rho$ and $r$ are related by 
\begin{equation}
\rho=2\ln\left(\frac{r}{r_{0}}\right)\,,\label{eq:ESolution_rho_r_relation}
\end{equation}
where we have set the value of the arbitrary scaling factor to be
$r_{0}$.

Now, defining the central pressure $p_{c}:=p\left(\rho\to-\infty\right)$,
from Eqs.~\eqref{eq:ESolution_constant_energy_solutions_p} and \eqref{eq:ESolution_constant_energy_solutions_boundary_P0},
we have

\begin{equation}
p_{c}=-\frac{\mu_{c}\left(1-\sqrt{1-\frac{2M}{r_{0}}}\right)}{1-3\sqrt{1-\frac{2M}{r_{0}}}}+\frac{\delta_{c}^{2}\left(1-2\sqrt{1-\frac{2M}{r_{0}}}\right)}{4\pi\left(1-3\sqrt{1-\frac{2M}{r_{0}}}\right)}\,,\label{eq:ESolution_constant_energy_central_pressure}
\end{equation}
where $\mu_{c}\equiv\mu\left(\rho\to-\infty\right)$ and $\delta_{c}\equiv\delta\left(\rho\to-\infty\right)$.
If we compared directly the above expression to a similar system in
GR, we would see that the second term on the right-hand side of Eq.~\eqref{eq:ESolution_constant_energy_central_pressure}
represents an explicit contribution due to the presence of spin. However,
there is a subtlety: Eq.~\eqref{eq:ESolution_constant_energy_Sch_mass}
indicates that the presence of spin also modifies the matching radius
$r_{0}$ and the value of the Schwarzschild parameter $M$, making
it difficult to draw conclusions only on \eqref{eq:ESolution_constant_energy_central_pressure}.

A clearer idea of the differences between our case and GR can be obtained
by computing the maximum mass that can be held by a star with constant
radius. Considering Eq.~\eqref{eq:ESolutions_constant_energy_assumption}
and if neither the densities $\mu_{c}$ and $\delta_{c}$ diverge,
the central pressure in Eq.~\eqref{eq:ESolution_constant_energy_central_pressure}
will go to infinity when $r_{0}=\frac{9}{4}M$ or, using Eq.~\eqref{eq:ESolution_constant_energy_Sch_mass},
when 
\begin{equation}
M_{\text{max}}=\frac{4}{9\sqrt{3\pi}}\left(\mu-\frac{\delta^{2}}{8\pi}\right)^{-\frac{1}{2}}\,.\label{eq:ESolution_constant_energy_Buchdahl_1}
\end{equation}
This results makes it clear that, when compared to a system with the
same energy density $\mu$ in GR, the presence of spin increases the
maximum allowed mass.

In analogy with the calculation of the Buchdahl limit in GR we can
generalize this discussion to non constant $\tilde{\mu}$. Consider
the quantity $\tilde{\mu}:=8\pi\mu-\delta^{2}$ and assume it to be
non-negative and $d\tilde{\mu}/dr\le0$, for $r\in\left[0,r_{0}\right]$.
Following the same reasoning of Ref.~\citep{Buchdahl} (see also
\citep{Wald}) we can find an upper limit for the amount of mass a
star with constant radius can hold: 
\begin{equation}
\frac{m\left(r_{0}\right)}{r_{0}}\leq\frac{4}{9}\,,\label{eq:ESolution_constant_energy_Buchdahl_2}
\end{equation}
with 
\begin{equation}
m\left(r_{0}\right)=\frac{1}{2}\int_{0}^{r_{0}}\tilde{\mu}\left(r\right)\,r^{2}dr\,.\label{eq:ESolution_constant_energy_Buchdahl_2_mass_definition}
\end{equation}

At first sight, the expression in Eq.~\eqref{eq:ESolution_constant_energy_Buchdahl_2}
matches the one found by Buchdahl \citep{Buchdahl} for GR. However,
there is a correction due to the presence of spin in the function
$m\left(r_{0}\right)$, Eq.~\eqref{eq:ESolution_constant_energy_Buchdahl_2_mass_definition},
leading us to conclude that for the same value of the circumferential
radius, $r_{0}$, a star can hold more matter in the presence of spin
than in the null-spin case. It is also worth mentioning that the quantity
$m\left(r_{0}\right)$ agrees with the value of the Schwarzschild
parameter of the exterior space-time, therefore, the gravitational
mass of such objects is determined not only by the energy density,
$\mu$, but also by the spin density, $\delta$, which was expected
because of the specific way in which spin gravitates in our specific
Einstein-Cartan setup.

As a final comment, although \emph{a priori} there is nothing that
forces $\delta^{2}$ to be smaller than $8\pi\mu$, it is expected
that in stars - even neutron stars - $\delta^{2}\ll\mu$ (see Refs.~\citep{Kerlick,Hehl2}),
hence, $\tilde{\mu}\geq0$, as was assumed in the derivation of Eq.~\eqref{eq:ESolution_constant_energy_Buchdahl_2}.
On the other hand, the requirement that $d\tilde{\mu}/dr\le0$ might
not be as physically reasonable as in the case of GR since, as we
will see bellow, the presence of spin allows for a richer possible
behavior for the matter variables.

\subsection{Spin held stars\label{subsec:Spin-held-stars}}

In the previous subsection we have considered a classical model for
a relativistic star which is similar to the simplest model for this
type of objects in General Relativity. However the presence of spin
allows solutions which are not contemplated in the Einstein's theory.
The prototype of such objects is a star which is supported only by
the gravitation of the spin of the Weyssenhoff fluid. In the remaining
of the subsection, we will analyze this case and prove the following
result 
\begin{prop}
\label{prop:no_spin_held_stars}There are no static, spherically symmetric
solutions of the Einstein-Cartan theory sourced by a Weyssenhoff fluid with null isotropic pressure
that have all the following properties 
\begin{enumerate}
\item $\delta\left(r\right)$ is non-null for $r\in\left[0,r_{0}\right[$
and $\delta\left(r_{0}\right)=0$, for some $r_{0}>0$; 
\item $\delta^{2}\left(r\right)$ is a monotonically decreasing function
for all $r\in\left[0,r_{0}\right]$; 
\item the spin and energy density functions: $\delta\left(r\right)$ and
$\mu\left(r\right)$, are at least of class $C^{1}$ and the function
$\mathcal{A}\left(r\right)$ is differentiable for all $r\in\left[0,r_{0}\right]$; 
\item the function $M\left(r\right):=\frac{1}{2}\int_{0}^{r}\left[8\pi\,\mu\left(r\right)-\delta^{2}\left(r\right)\right]x^{2}dx$
is such that $2M\left(r\right)<r$, for all $r\in\left]0,r_{0}\right]$. 
\end{enumerate}
\end{prop}

To prove Proposition \ref{prop:no_spin_held_stars} we will consider
first the behavior of the quantities of interest in a neighborhood
of the center, $r=0$, and then on the boundary of the star. In doing
so, in order to make the reasoning more intuitive, we shall consider
here that the integral curves of the vector field $e$ are parameterized
by the circumferential radius $r$.

Defining the quantities 
\begin{equation}
\begin{aligned}\tilde{\mu}\left(r\right) & =8\pi\,\mu\left(r\right)-\delta^{2}\left(r\right)\,,\\
\tilde{p}\left(r\right) & =8\pi\,p\left(r\right)-\delta^{2}\left(r\right)\,,
\end{aligned}
\end{equation}
we find from the structure equations 
\begin{align}
\frac{r}{2}\phi\left(r\right)\tilde{p}_{,r} & =-\mathcal{A}\left(\tilde{\mu}+\tilde{p}\right)\,,\label{eq:Momentum_conservation}\\
\frac{r}{2}\phi\left(r\right)\mathcal{A}_{,r} & +\mathcal{A}^{2}+\mathcal{A}\phi=\frac{1}{2}\left(\tilde{\mu}+3\tilde{p}\right)\,,\label{eq:Propagation_A}\\
\tilde{p} & =\mathcal{A}\phi-K+\frac{1}{4}\phi^{2}\,,\label{eq:Constraint_p_A}
\end{align}
with 
\begin{align}
K\left(r\right) & =\frac{1}{r^{2}}\,,\\
\phi\left(r\right) & =\frac{2}{r}\sqrt{1-\frac{2M\left(r\right)}{r}}\,,\label{eq:phi_exact_solution}
\end{align}
and 
\begin{equation}
M\left(r\right)=\frac{1}{2}\int_{0}^{r}\tilde{\mu}\left(x\right)x^{2}dx\,,\label{eq:Mass_function}
\end{equation}
where, without loss of generality, we chose the direction of $e$
so that $\phi\left(r\right)$ is non-negative. Moreover, from Eqs.~\eqref{eq:Constraint_p_A}
and \eqref{eq:phi_exact_solution} we find the useful relation
\begin{equation}
\mathcal{A}\phi=\frac{2M\left(r\right)}{r^{3}}+\tilde{p}\,.\label{eq:Constraint_p_A_alt}
\end{equation}

We will consider now the case of a static, spherically symmetric compact
object held entirely by spin, that is, the case when $p\left(r\right)=0$
and $\tilde{p}\left(r\right)=-\delta^{2}\left(r\right)$, smoothly
matched to an exterior space-time modeled by a vacuum solution of
the Einstein-Cartan field equations. Moreover, we will assume that
for $r>0$, $2M\left(r\right)<r$, otherwise the scalar $\phi\left(r\right)$
would take complex values.

\subsubsection{Behavior at the center}

Assuming that the functions $\mu\left(r\right)$, $\delta^{2}\left(r\right)\in C^{1}$
we can write in a small enough neighborhood of $r=0$: 
\begin{equation}
\begin{aligned}\mu\left(r\right) & =\mu\left(0\right)+\mu_{,r}\left(0\right)\,r\,,\\
\delta^{2}\left(r\right) & =\delta^{2}\left(0\right)+\left(\delta^{2}\right)_{,r}\left(0\right)\,r\,,
\end{aligned}
\label{eq:Center_densities_expansion_at_0}
\end{equation}
where comma represents partial - or total - derivative with respect
to the variable in front. From Eqs.~\eqref{eq:Center_densities_expansion_at_0},
we find that in a small enough neighborhood of $r=0$, the mass function
\eqref{eq:Mass_function} is described by 
\begin{equation}
2M\left(r\right)=\frac{\tilde{\mu}\left(0\right)}{3}r^{3}+\frac{1}{4}\left(\tilde{\mu}_{,r}\left(0\right)\right)r^{4}\,.\label{eq:Center_mass_expansion}
\end{equation}
In particular we find that $M\left(r\right)$ goes to zero at least
as fast as $r^{3}$.

Now, Eqs.~\eqref{eq:Momentum_conservation} and \eqref{eq:Constraint_p_A_alt}
yield 
\begin{equation}
\frac{2}{r}\left(1-\frac{2M\left(r\right)}{r}\right)\frac{d\delta^{2}}{dr}=\left(\frac{2M\left(r\right)}{r^{3}}-\delta^{2}\right)\left(\tilde{\mu}-\delta^{2}\right)\,.\label{eq:Conservation_momentum_alt}
\end{equation}
In a region where $r\in\left[0,\epsilon\right[$, with $\epsilon\ll1$,
the RHS of this equation takes values in $\mathbb{R}$, therefore
\begin{equation}
\left(\delta^{2}\right)_{,r}\left(0\right)=0\,.\label{eq:Center_derivative_delta_at_0}
\end{equation}
Repeating the same reasoning in Eq.~\eqref{eq:Constraint_p_A_alt}
we find that

\begin{equation}
\mathcal{A}\left(0\right)=0\,.\label{eq:Center_A_at_0}
\end{equation}

Let us now assume that there exists an $r_{a}>0$ where for $r\in\left]0,r_{a}\right[$,
$\mathcal{\mathcal{A}}\left(r\right)>0$. From Eq.~\eqref{eq:Momentum_conservation}
we will find that in this region $\tilde{\mu}+\tilde{p}\leq0$ which
implies that $\tilde{\mu}+3\tilde{p}<0$. Then, from Eq.~\eqref{eq:Propagation_A}
we find that $\mathcal{A}_{,r}\left(r\right)<0$, for all $r\in\left]0,r_{a}\right[$.
This, however, violates the initial hypothesis since, $\mathcal{A}\left(0\right)=0$
and we assume that $\mathcal{A}\left(r\right)>0$, for $r\in\left]0,r_{a}\right[$,
that is, $\mathcal{A}_{,r}\left(r\right)$ would have to be positive
for some $r\in\left[0,r_{a}\right[$ .

Another possibility is that for a region $r\in\left[0,r_{a}\right]$,
$\mathcal{A}\left(r\right)=0$ and for $r\in\left]r_{a},r_{c}\right]$
with $r_{c}>r_{a}$, $\mathcal{A}\left(r\right)>0$. If this were
the case, since for $r\in\left]r_{a},r_{c}\right]$, $\mathcal{A}\left(r\right)>0$
, there would exist a value $r_{b}\in\left]r_{a},r_{c}\right]$ such
that $\mathcal{A}_{,r}\left(r_{b}\right)>0$. Using this in Eq.~\eqref{eq:Propagation_A},
at $r=r_{b}$ we find 
\begin{equation}
\tilde{\mu}+3\tilde{p}>0\Rightarrow\tilde{\mu}+\tilde{p}>0\,,\label{eq:Center_ineq_1}
\end{equation}
but from Eq.~\eqref{eq:Momentum_conservation} and imposing that
$\left(\delta^{2}\right)_{,r}\le0$ we find that: $\left.\tilde{\mu}+\tilde{p}\right|_{r=r_{b}}\leq0$,
contradicting \eqref{eq:Center_ineq_1}.

Another possibility is that $\mathcal{A}\left(r\right)=0$, for all
$r\in\left[0,r_{0}\right]$. From Eqs.~\eqref{eq:Momentum_conservation}
and \eqref{eq:Propagation_A} this simply represents a vacuum solution
as such it does not represent a solution for a compact object.

Gathering this results we conclude that there exists an $r_{d}>0$
such that in the region $\left[0,r_{d}\right[$, $\mathcal{A}\left(r\right)\leq0$
and it must take negative values in some sub-region.

\subsubsection{Behavior at the boundary}

Let us now define the boundary of the compact object as the hypersurface
at which the spin density goes to zero, that is, $\delta^{2}\left(r_{0}\right)=0$.
In such hypersurface we have three possible behaviors for the function
$\mathcal{A}$: 
\begin{enumerate}
\item $\mathcal{A}\left(r_{0}\right)<0$; 
\item $\mathcal{A}\left(r_{0}\right)>0$; 
\item $\mathcal{A}\left(r_{0}\right)=0$. 
\end{enumerate}
Let us consider each case separately.

\paragraph{(1)The case $\mathcal{A}\left(r_{0}\right)<0$\label{paragraph:SpinHeld_A_r0_negative}
\protect \protect \\
 }

\noindent From Eq.~\eqref{eq:Constraint_p_A_alt} we have that at
$r=r_{0}$ 
\begin{equation}
\left.\mathcal{A}\phi\right|_{r=r_{0}}=\frac{2M\left(r_{0}\right)}{r_{0}^{3}}<0\,.
\end{equation}
Therefore, from Eq.~\eqref{eq:Mass_function} there exists a region
$\left]r_{f},r_{g}\right[$ where 
\begin{equation}
\tilde{\mu}\left(r\right)<0\,,\label{eq:Boundary_case_1_constraint_mu_1}
\end{equation}
then $\tilde{\mu}+\tilde{p}<0$, in that region. From Eq.~\eqref{eq:Momentum_conservation},
to guarantee that the spin density is a monotonically decreasing function
of $r$, we find that $\mathcal{A}\left(r\right)\geq0$, for $r\in\left]r_{f},r_{g}\right[$.
So, either $\mathcal{A}\left(r\right)=0\,\wedge\,\mathcal{A}_{,r}\left(r\right)=0$
for all $r\in\left]r_{f},r_{g}\right[$, that is, the function $\mathcal{A}\left(r\right)$
takes the value zero and stays zero for all $r\in\left]r_{f},r_{g}\right[$;
or $\mathcal{A}\left(r\right)>0$ for some $r\in\left]r_{f},r_{g}\right[$.
The former case is not possible: from Eq.~\eqref{eq:Propagation_A},
$\tilde{\mu}\left(r\right)+3\tilde{p}\left(r\right)=0$, hence, $\tilde{\mu}\left(r\right)\geq0$,
for all $r\in\left]r_{f},r_{g}\right[$, which contradicts the inequality
\eqref{eq:Boundary_case_1_constraint_mu_1}. As for the latter - the
case when $\mathcal{A}\left(r\right)>0$, for some $r\in\left]r_{f},r_{g}\right[$
- in the previous sub-section it was shown that for some sub-region
of $\left[0,r_{d}\right[$, $\mathcal{A}\left(r\right)\leq0$, therefore
the region $\left]r_{f},r_{g}\right[$ cannot be a sub-region of $\left[0,r_{d}\right[$.
With this said, since $\mathcal{A}\left(r\right)$ is a differentiable
function, there exists a region with, say, $r=r_{e}<r_{0}$, where
$\mathcal{A}\left(r_{e}\right)>0\,\wedge\,\mathcal{A}_{,r}\left(r_{e}\right)>0$.
Then, from Eq.~\eqref{eq:Propagation_A} 
\begin{equation}
\left.\tilde{\mu}+3\tilde{p}\right|_{r=r_{e}}>0\Rightarrow\left.\tilde{\mu}+\tilde{p}\right|_{r=r_{e}}>0\,.\label{eq:Boundary_case_1_constraint_mu_2}
\end{equation}
However, substituting this results in Eq.~\eqref{eq:Momentum_conservation}
we find: $\left(\delta^{2}\right)_{,r}\left(r_{e}\right)>0$, which
contradicts the assumption that the spin density is a monotonically
decreasing function.

\paragraph{(2) The case $\mathcal{A}\left(r_{0}\right)>0$ \protect \protect \\
 }

\noindent For the case when $\mathcal{A}\left(r_{0}\right)>0$, we
can simply repeat the proof in the previous sub-subsection and conclude
in the same way that the assumptions are violated in a region. We
just remark that the point with radial coordinate $r=r_{e}$, in the
proof, can always be chosen such that $r_{e}<r_{0}$ since, for whatever
the value of $\mathcal{A}\left(r_{0}\right)>0$, there is a point
where $0<\mathcal{A}\left(r<r_{0}\right)<\mathcal{A}\left(r_{0}\right)$.

\paragraph{(3)The case $\mathcal{A}\left(r_{0}\right)=0$ \protect \protect \\
 }

\noindent In this the case when $\mathcal{A}\left(r_{0}\right)=0$
we have, from Eq.~\eqref{eq:Constraint_p_A_alt} that 
\begin{equation}
M\left(r_{0}\right)=0\,.\label{eq:Boundary_zero_mass_at_r0}
\end{equation}
From this we have three possibilities:
\begin{enumerate}[label=(\alph*)]
\item $\tilde{\mu}\left(r\right)=0$, for $r\in\left[0,r_{a}\right]$; 
\item $\tilde{\mu}\left(r\right)<0$, for $r\in\left]0,r_{a}\right]$; 
\item $\tilde{\mu}\left(r\right)>0$, for $r\in\left]0,r_{a}\right]$; 
\end{enumerate}
for some $r_{a}>0$.

Let us consider each case individually.

(a) In the case when $\tilde{\mu}\left(r\right)=0$, for $r\in\left[0,r_{a}\right]$
we have from Eq.~\eqref{eq:Constraint_p_A_alt} that $\mathcal{A}\left(r\right)<0$.
However, using this result in \eqref{eq:Momentum_conservation} we
see that it implies that the spin density is an increasing function
of $r$, violating the hypothesis.

(b) In the case when the corrected energy density is such that $\tilde{\mu}\left(r\right)<0$,
for $r\in\left]0,r_{a}\right]$, from \eqref{eq:Mass_function} we
have that the mass function is negative, in this region. From Eq.~\eqref{eq:Constraint_p_A_alt}
we than conclude that $\mathcal{A}\left(r\right)<0$, $r\in\left]0,r_{a}\right]$.
However, going back to Eq.~\eqref{eq:Momentum_conservation} we find
that the spin density is an increasing function of $r$, violating
the hypothesis.

(c) Finally, consider the case when $\tilde{\mu}\left(r\right)>0$,
for $r\in\left]0,r_{a}\right]$. From \eqref{eq:Mass_function}, this
implies that the mass function is positive in this region. Since Eq.~\eqref{eq:Boundary_zero_mass_at_r0}
must be verified, there must be a region where $\tilde{\mu}\left(r\right)<0$.
We can then repeat the arguments of the case $\mathcal{A}\left(r_{0}\right)<0$,
which lead to the conclusion that the hypothesis would be violated
in some region inside the star.

Gathering the previous results we have proven the result in Proposition
\ref{prop:no_spin_held_stars}.

We end this Section by remarking that if instead of imposing $\tilde{p}=-\delta^{2}$
we only imposed that $p<\delta^{2}$, that is, the thermodynamical
pressure is always smaller than correction due to the spin density,
then all the previous results are valid if $\tilde{p}_{,r}\ge0$.
Notice, however, that in this scenario this condition, simply measures
the gradient of the quantity $8\pi p-\delta^{2}$.

\subsection{Reconstructing exact solutions}

As in the case of the theory of General Relativity, when torsion is
present it is possible to generate exact solutions via reconstruction
algorithms \citep{Sante1,Sante2}. The idea is to assign a given metric
tensor and deduce the corresponding behavior of the energy density,
pressure and spin density.

Analyzing Eqs.~\eqref{eq:Alt_SE_prop_Y} - \eqref{eq:Alt_SE_prop_K}
and using Eqs.~\eqref{eq:Alt_SE_constraint_MPYEO} and \eqref{eq:Alt_SE_constraint_YOTPK},
shows that differently from the case of anisotropic compact objects
in General Relativity \citep{Sante2}, the structure equations cannot
be solved for the spin density. This implies that the reconstruction
algorithm can only be used if an additional relation is provided,
either relating the spin density to the other matter variables or
an equation of state for matter.

In the following we will show some applications of this algorithms
which return some interesting solutions from a physical point of view.

\subsubsection{Connecting the spin density to the energy density: ``Buchdhal stars''}

A natural additional relation is to have the spin density to be proportional
to the energy density of the Weissenhoff fluid. In this case, however,
the junction conditions that we have seen in Sec.~\ref{1+1+2Junction}
pose the problem to have both the energy density and the pressure
to be zero at the boundary. A class of solutions which are devised
to have exactly this property was given by Buchdhal \citep{Buch3}.
We will now reconstruct this solutions in the case of equations \eqref{eq:TOV_static_P}
- \eqref{eq:TOV_static_E}.

Consider a spherically symmetric space-time characterized by the line
element 
\begin{equation}
ds^{2}=-A\left(w\right)dt^{2}+B\left(w\right)dw^{2}+C\left(w\right)\big(d\theta^{2}+\sin^{2}\theta\,d\varphi^{2}\big)\,,\label{eq:Rec1_line_element_sph_sym_general}
\end{equation}
where 
\begin{equation}
\begin{aligned}\eta\left(w\right) & =\frac{(a-1)\sin\left(Rw\right)}{Rw}\,, &  &  & A\left(w\right) & =\frac{a\left(1+a-\eta\right)}{1+a+\eta}\,,\\
B\left(w\right) & =\frac{(1+a+\eta)}{a(1+a-\eta)} &  &  & C\left(w\right) & =\frac{w^{2}(1+a+\eta)^{2}}{4a^{2}}\,,
\end{aligned}
\label{eq:Rec1_line_element_Buch3}
\end{equation}
and $w$ is connected to $\rho$ by the relation 
\begin{equation}
e^{\rho}=\frac{w^{2}}{4a^{2}}\left(1+a+\eta\right)^{2}\,.\label{eq:Rec1_w_rho_relation}
\end{equation}
Notice that the circumferential radius, $r$, vanishes when $w=0$.

From Eqs.~\eqref{eq:Alt_SE_prop_Y}, \eqref{eq:Alt_SE_prop_K}, \eqref{eq:TOV_static_K}
and \eqref{eq:TOV_static_Y}, assuming $\mathbb{W}=\Delta$ and $\Delta^{2}=\gamma\mathcal{M}$,
we find

\begin{align}
\mathcal{M} & =\frac{2\mathcal{K}_{,\rho}+4\mathcal{K}^{2}-\mathcal{K}}{4(1-\gamma)\mathcal{K}}\,,\label{eq:Rec1_MDM}\\
\mathcal{P} & =Y-\mathcal{K}+\frac{1}{4}-\frac{2\mathcal{K}_{,\rho}+4\mathcal{K}^{2}-\mathcal{K}}{4(1-\gamma)\mathcal{K}}\,,\label{eq:Rec1_PDM}\\
0 & =(2Y+1)\mathcal{K}_{,\rho}-4\mathcal{K}^{2}-\mathcal{K}\left[4Y_{,\rho}+4(Y-1)Y-1\right]\,.\label{eq:Rec1_constr_DgM}
\end{align}
The form of $Y$ and $\mathcal{K}$ that satisfies the constraint
\eqref{eq:Rec1_constr_DgM} can be found directly from their definition
in a general coordinate system (see Refs.~\citep{Sante1,Sante2})
\begin{equation}
\begin{aligned}Y & =\frac{1}{2}\frac{CA_{,w}}{AC_{,w}}=\frac{\left(1+a\right)w\eta_{,w}}{2\left(\eta-a-1\right)\left(1+a+\eta+w\eta_{,w}\right)}\,,\\
\mathcal{K} & =\frac{BC}{\left(C_{,w}\right)^{2}}=\frac{a\left(1+a+\eta\right)}{\left(1+a-\eta\right)\left(1+a+\eta+w\eta_{,w}\right)^{2}}\,.
\end{aligned}
\label{eq:Rec1_Y_K}
\end{equation}
Using Eqs.~\eqref{eq:Rec1_MDM}, \eqref{eq:Rec1_PDM} and \eqref{eq:Rec1_Y_K},
the energy density and the pressure are then given by 
\begin{equation}
\begin{aligned}\mu & =\frac{aR^{2}\eta\left(3\eta-2-2a\right)}{8\pi\left(\gamma-1\right)\left(1+a+\eta\right)^{2}}\,,\\
p & =\frac{aR^{2}\eta\left[2\gamma\left(2\eta-a-1\right)-\eta\right]}{8\pi\left(\gamma-1\right)(1+a+\eta)^{2}}\,.
\end{aligned}
\label{eq:Rec1_matter_variables}
\end{equation}

\begin{figure}
\subfloat[Coefficients of the metric in Eqs.~\eqref{eq:Rec1_line_element_sph_sym_general}
and \eqref{eq:Rec1_line_element_Buch3}.]{\includegraphics[width=0.9\columnwidth]{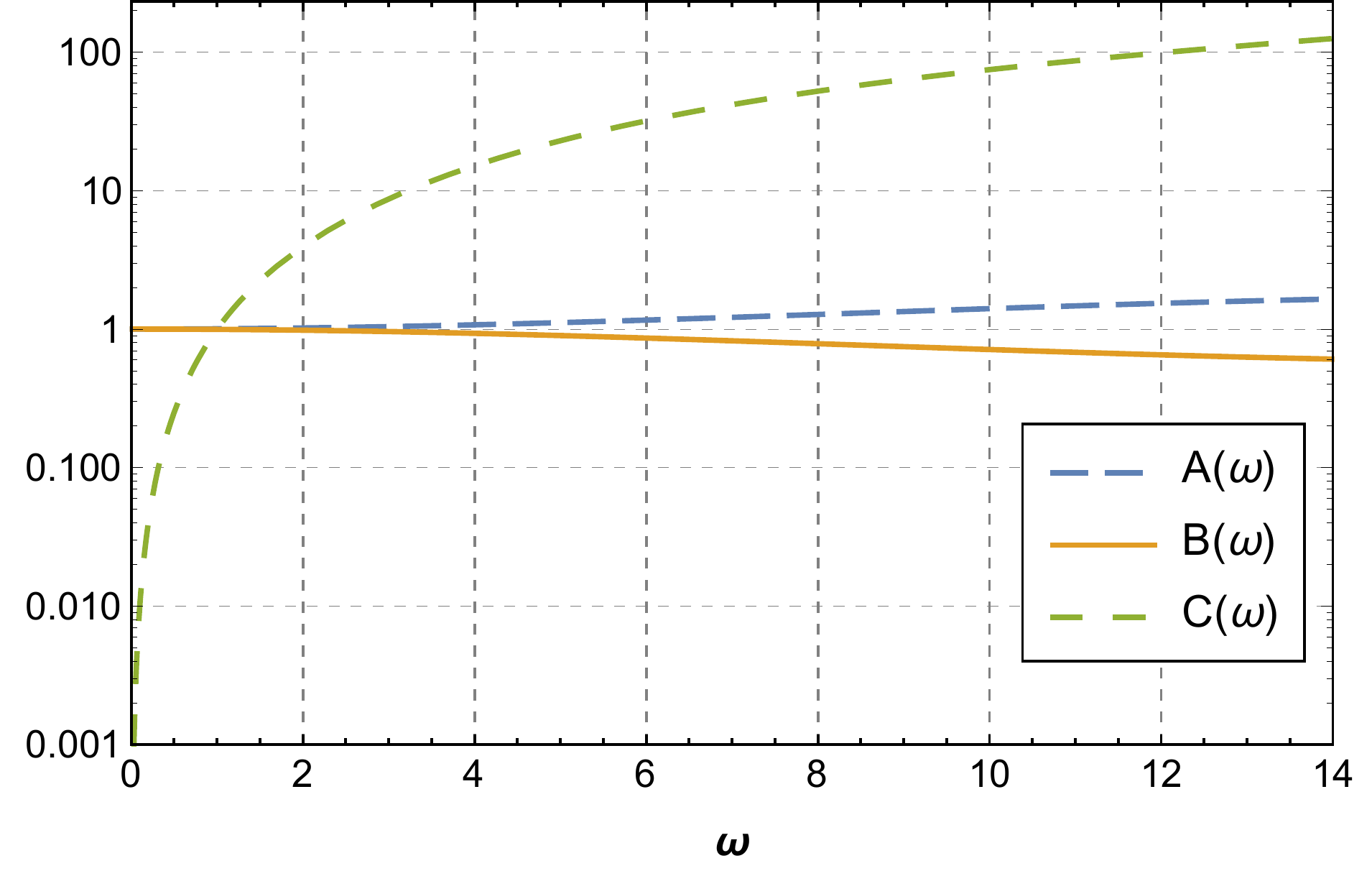}

}\\
\subfloat[Thermodynamic quantities in Eqs.~\eqref{eq:Rec1_matter_variables}
and the spin density.]{\hspace*{-0.4cm}\includegraphics[width=0.92\columnwidth]{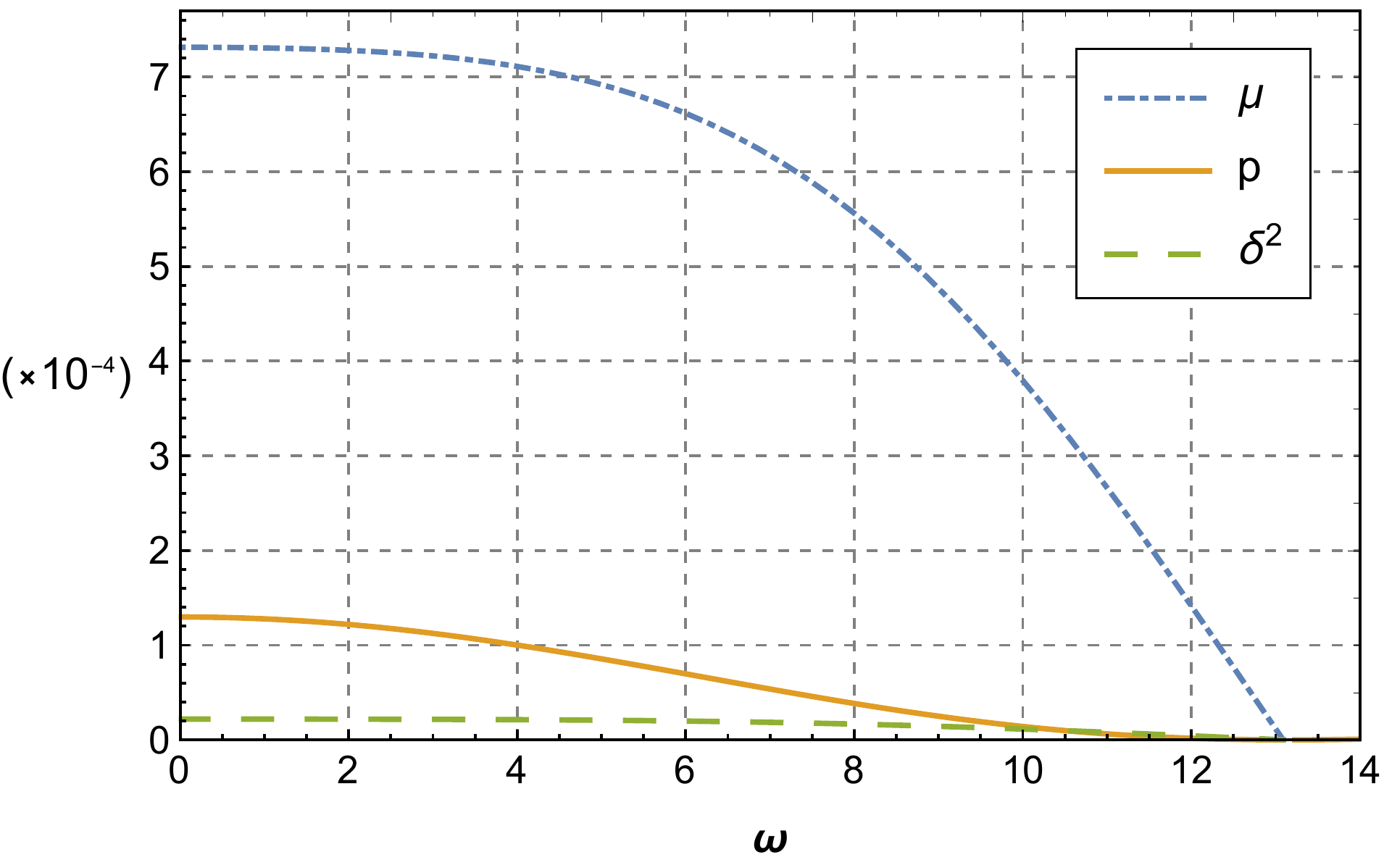}

}

\caption{\label{Fig:Rec1_1} Plots of the behavior of the metric components,
(a) and matter variables (b) associated with the solution in Eqs.~\eqref{eq:Rec1_line_element_sph_sym_general}
and \eqref{eq:Rec1_line_element_Buch3} for $a=1.6$, $\gamma=0.03/(8\pi)$
and $R=0.24$.}

\end{figure}

\begin{figure}
\subfloat[Coefficients of the metric in Eqs.~\eqref{eq:Rec1_line_element_sph_sym_general}
and \eqref{eq:Rec1_line_element_Buch3}.]{\includegraphics[width=0.9\columnwidth]{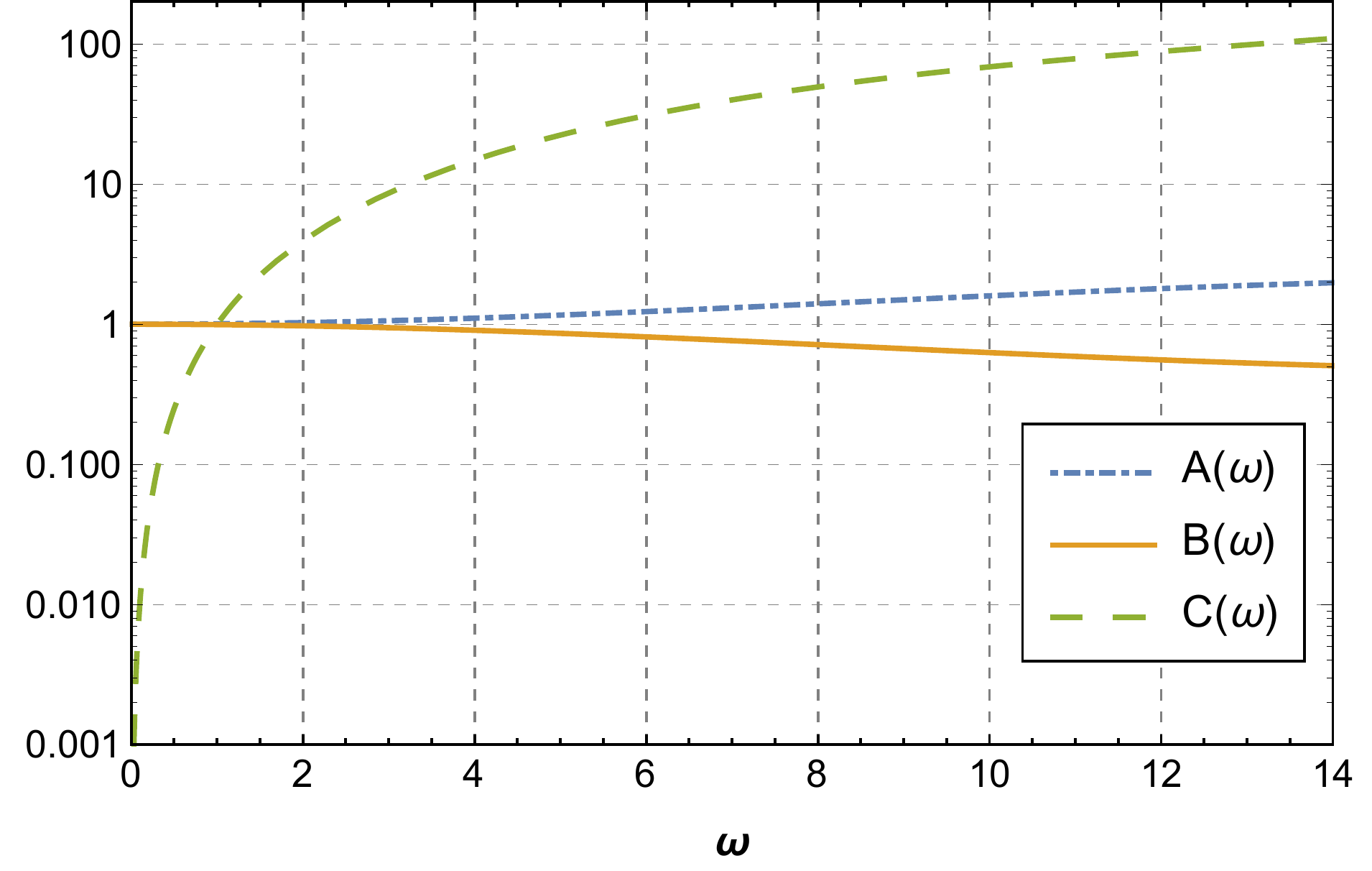}

}\\
\subfloat[Thermodynamic quantities in Eqs.~\eqref{eq:Rec1_matter_variables}
and the spin density.]{\hspace*{-0.4cm}\includegraphics[width=0.92\columnwidth]{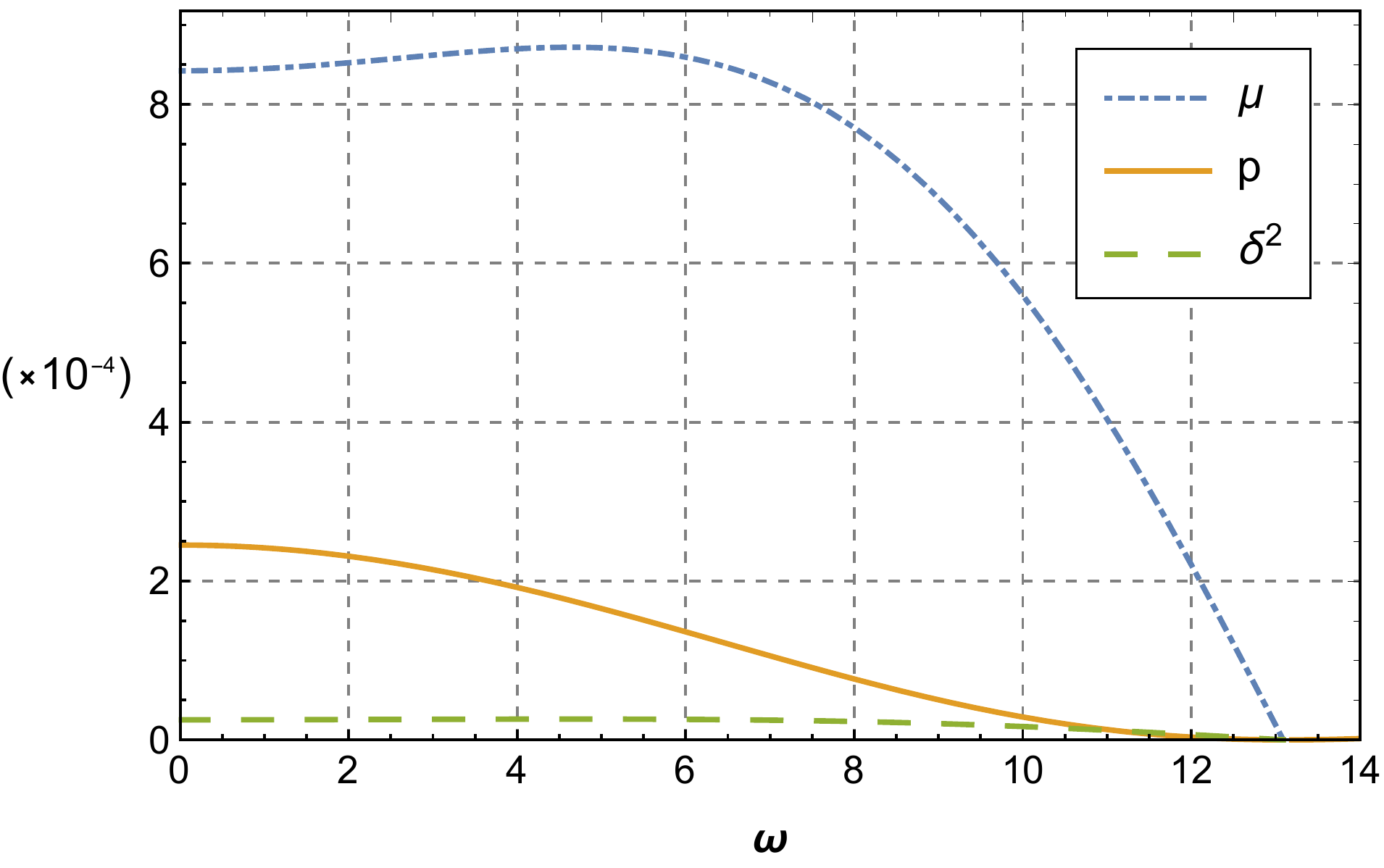}

}

\caption{\label{Fig:Rec1_2} Plots of the behavior of the metric components,
(a) and matter variables (b) associated with the solution in Eqs.~\eqref{eq:Rec1_line_element_sph_sym_general}
and \eqref{eq:Rec1_line_element_Buch3} for $a=1.9$, $\gamma=0.03/(8\pi)$
and $R=0.24$.}
\end{figure}

\begin{figure}
\subfloat[Coefficients of the metric in Eqs.~\eqref{eq:Rec1_line_element_sph_sym_general}
and \eqref{eq:Rec1_line_element_Buch3}.]{\includegraphics[width=0.9\columnwidth]{Recon1_metric_a_1d9_gamma_d5_R_d24}

}\\
\subfloat[Thermodynamic quantities in Eqs.~\eqref{eq:Rec1_matter_variables}
and the spin density.]{\hspace*{-0.4cm}\includegraphics[width=0.92\columnwidth]{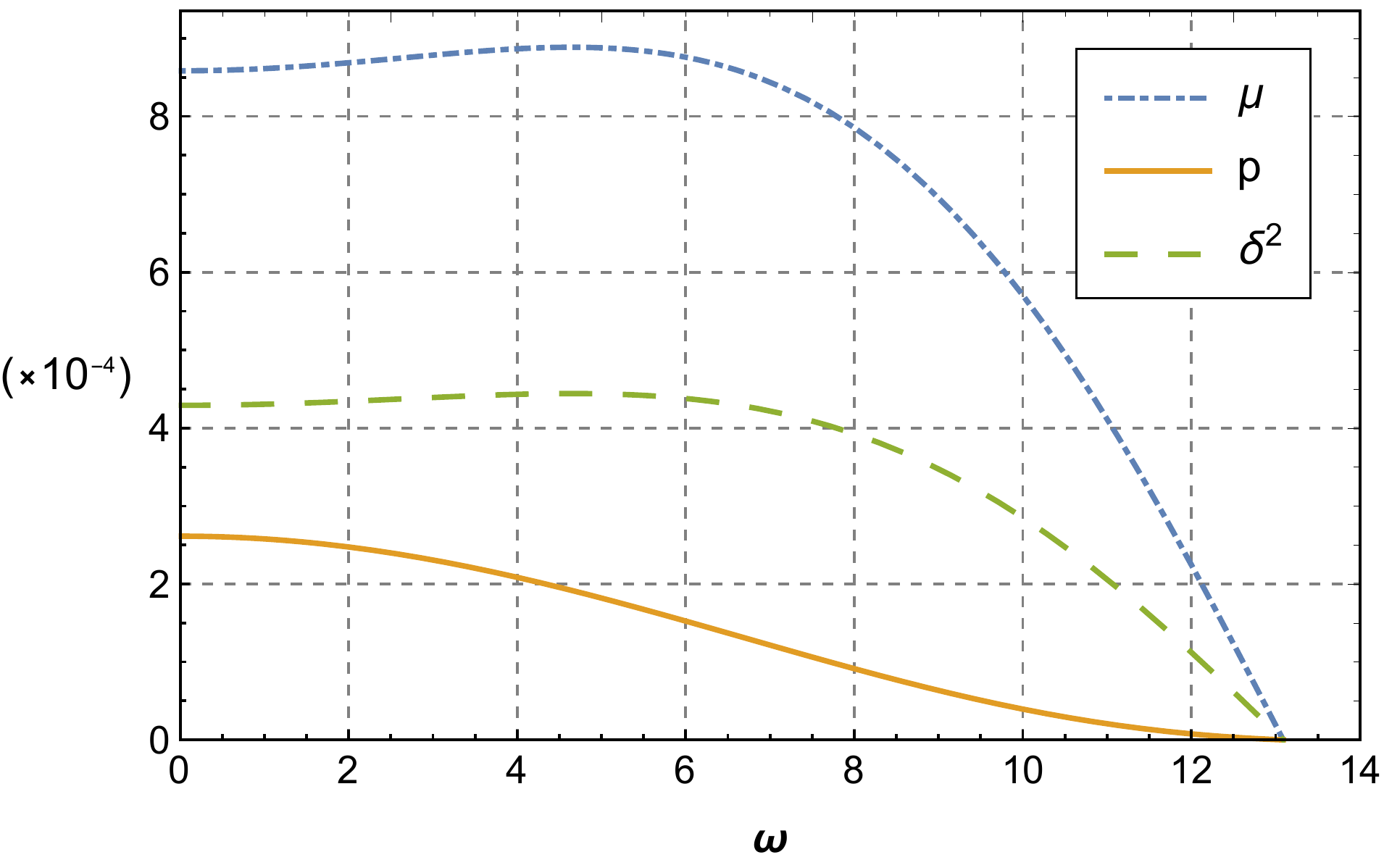}

}

\caption{\label{Fig:Rec1_3} Plots of the behavior of the metric components,
(a) and matter variables (b) associated with the solution in Eqs.~\eqref{eq:Rec1_line_element_sph_sym_general}
and \eqref{eq:Rec1_line_element_Buch3} for $a=1.9$, $\gamma=0.5/(8\pi)$
and $R=0.24$.}
\end{figure}

As said, this family of solutions have, by construction, the property
that the pressure, energy and spin densities all vanish at a particular
hypersurface. In Figures~\ref{Fig:Rec1_1} - \ref{Fig:Rec1_3} we
present the behavior of these quantities for a few combinations of
the parameters, showing that the values of the parameters $a$ and
$\gamma$ have a direct impact in the profile of the densities, whereas,
the parameter $R$ defines the value when the matter variables go
to zero. Moreover, from the plots it is clear that the presence of
spin markedly changes the type of behavior the matter may have. In
particular, for certain values of the parameters $a$ and $\gamma$
the functions $\mu$, $p$ or $\delta$ might not be monotonically
decreasing functions of the coordinate $w$.

\subsubsection{Connecting the spin density to the pressure}

Another option that reduces the number of conditions related to the
junction is to associate the spin density to the pressure. This choice,
which at first might appear unnatural, corresponds to the case in
which the spin depends on the equation of state. We can imagine that
particles with spin will create different structures
not unlikely to the ones that characterize the crystalline phases
of water ice (see e.g. Ref.~\citep{ICE}). Our ansatz refers to this
kind of effects.

The reconstruction equations in this case, setting $\Delta^{2}=\gamma\mathcal{P}$,
read 
\begin{align}
\mathcal{M}= & \frac{4\mathcal{K}^{2}+2\mathcal{K}_{,\rho}-\mathcal{K}}{4\left(1-\gamma\right)\mathcal{K}}-\frac{\gamma\left[\mathcal{K}_{,\rho}+\mathcal{K}\left(4\mathcal{K}-2Y-1\right)\right]}{2\left(1-\gamma\right)\mathcal{K}}\,,\\
\mathcal{P}= & \frac{1-4\mathcal{K}+4Y}{4\left(1-\gamma\right)}\,,\\
0= & \left(2Y+1\right)\mathcal{K}_{,\rho}-4\mathcal{K}^{2}-\mathcal{K}\left[4Y_{,\rho}+4\left(Y-1\right)Y-1\right]\,.\label{eq:Rec2_constr_DgM}
\end{align}

Let us now consider a metric in which the $(0,0)$ coefficient, $A$,
is given by 
\begin{equation}
A=A_{0}\left(a+br_{0}^{2}e^{\rho}\right)^{2}\,,
\end{equation}
where $a,b$ and $r_{0}$ are arbitrary constants. From the definition
of $Y$ one obtains 
\begin{equation}
Y=\frac{1}{2}\frac{A_{,\rho}}{A}=\frac{br_{0}^{2}e^{\rho}}{a+br_{0}^{2}e^{\rho}}\,,
\end{equation}
and from Eq.~\eqref{eq:Rec2_constr_DgM} it follows that 
\begin{equation}
\mathcal{K}=\frac{\left(a+3br_{0}^{2}e^{\rho}\right)^{2/3}}{\mathcal{K}_{0}e^{\rho}+4\left(a+3br_{0}^{2}e^{\rho}\right)^{2/3}}\,,
\end{equation}
where $\mathcal{K}_{0}$ is an integration constant. In terms of the
area radius $r$, this result corresponds to the line element 
\begin{equation}
ds^{2}=-A\left(r\right)dt^{2}+B\left(r\right)dr^{2}+r^{2}\,\big(d\theta^{2}+\sin^{2}\theta\,d\varphi^{2}\big),\label{eq:Rec2_line_element_sph_sym_general}
\end{equation}
with 
\begin{equation}
\begin{aligned}A\left(r\right) & =A_{0}\left(a+br^{2}\right)^{2},\\
B\left(r\right) & =\left(1+\frac{c\,r^{2}}{\left(a+3br^{2}\right){}^{2/3}}\right)^{-1}\,.
\end{aligned}
\label{eq:Rec2_line_element_sph_sym_coef}
\end{equation}
The energy density and the pressure are given by 
\begin{equation}
\begin{aligned}\mu & =\frac{b^{2}r^{2}\left[5c\left(1-4\gamma\right)r^{2}-12\gamma\left(a+3br^{2}\right)^{2/3}\right]}{8\pi(\gamma-1)\left(a+br^{2}\right)\left(a+3br^{2}\right)^{5/3}}\\
 & \hspace{0.4cm}-\frac{4ab\left[\gamma\left(a+3br^{2}\right)^{2/3}+2c\left(2\gamma-1\right)r^{2}\right]}{8\pi(\gamma-1)\left(a+br^{2}\right)\left(a+3br^{2}\right)^{5/3}}+\\
 & \hspace{0.4cm}+\frac{a^{2}c\left(3-4\gamma\right)}{8\pi(\gamma-1)\left(a+br^{2}\right)\left(a+3br^{2}\right)^{5/3}}\,,\\
p & =\frac{4b\left(a+3br^{2}\right)^{2/3}+ac+5bcr^{2}}{8\pi(1-\gamma)\left(a+br^{2}\right)\left(a+3br^{2}\right)^{2/3}}\,.
\end{aligned}
\label{eq:Rec2_matter_variables}
\end{equation}

\begin{figure}[!ht]
\subfloat[Coefficients of the metric in Eqs.~\eqref{eq:Rec2_line_element_sph_sym_general}
and \eqref{eq:Rec2_line_element_sph_sym_coef}.]{\includegraphics[width=0.9\columnwidth]{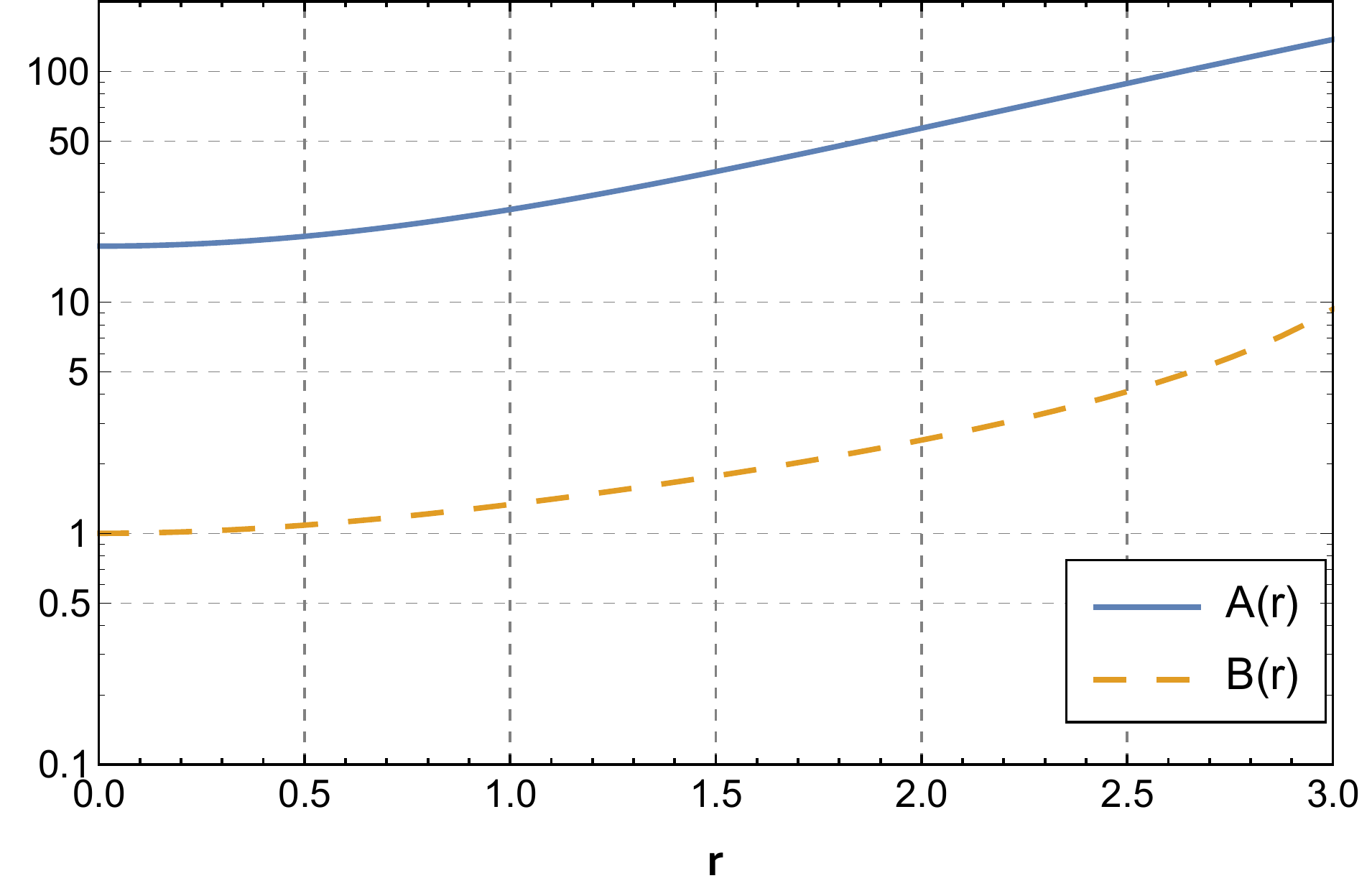}

}\\
\subfloat[The thermodynamic quantities in Eqs.~\eqref{eq:Rec2_matter_variables}.]{\hspace*{-0.45cm}\includegraphics[width=0.94\columnwidth]{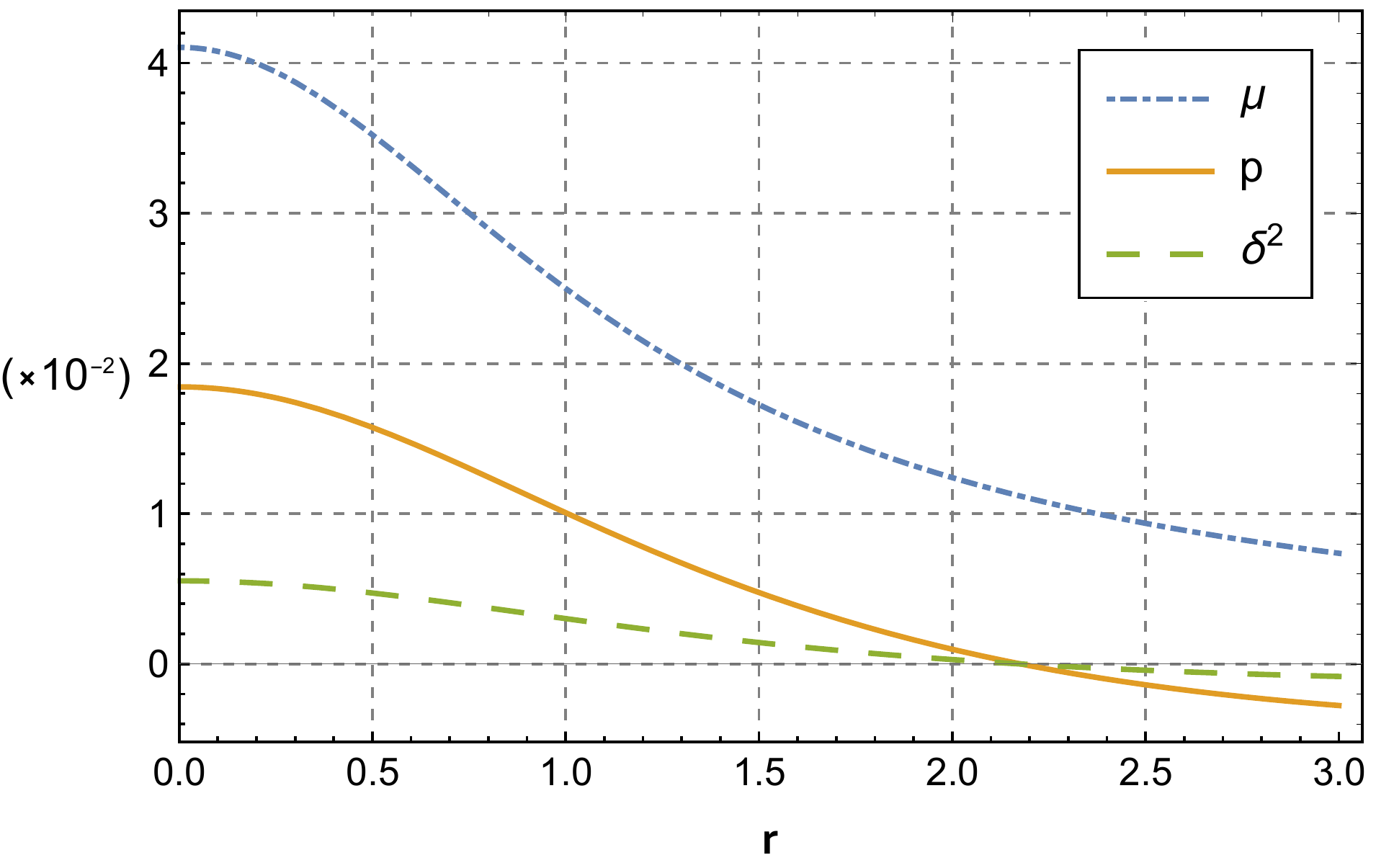}

}\caption{\label{Fig:Rec2} Plots of the behavior of the metric components,
(a) and matter variables (b) associated with the solution in Eqs.~\eqref{eq:Rec2_line_element_sph_sym_general}
- \eqref{eq:Rec2_matter_variables} in the case $a=5$, $b=1$, $c=-1$,
$\gamma=0.3/(8\pi)$ and $A_{0}=0.7$.}
\end{figure}

We give in Figure \ref{Fig:Rec2} the behavior of this solution for
specific values of the parameters, showing the existence of an hypersurface
where both $p$ and $\delta^{2}$ vanish, so that we can smoothly
match such solution with a vacuum exterior space-time.

\begin{figure}[!ht]
\subfloat[The coefficients of the metric in Eq.~\eqref{eq:Rec2_line_element_sph_sym_general}
and \eqref{eq:Rec3_line_element_sph_sym_coef}.]{\includegraphics[width=0.88\columnwidth]{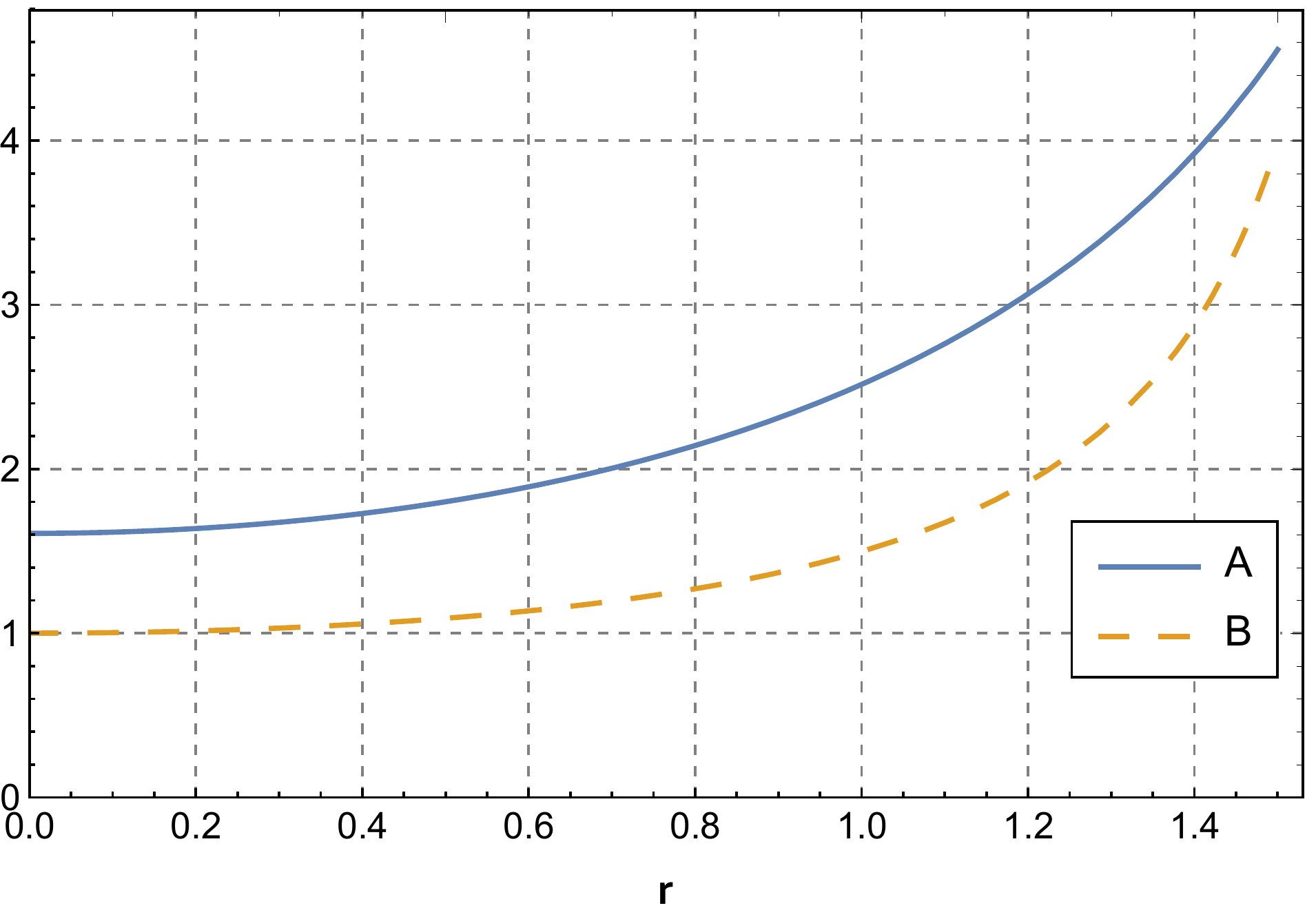}

}\\
\subfloat[The thermodynamic quantities in Eqs.~\eqref{eq:Rec3_matter_variables}.]{\hspace*{-0.8cm}\includegraphics[width=0.98\columnwidth]{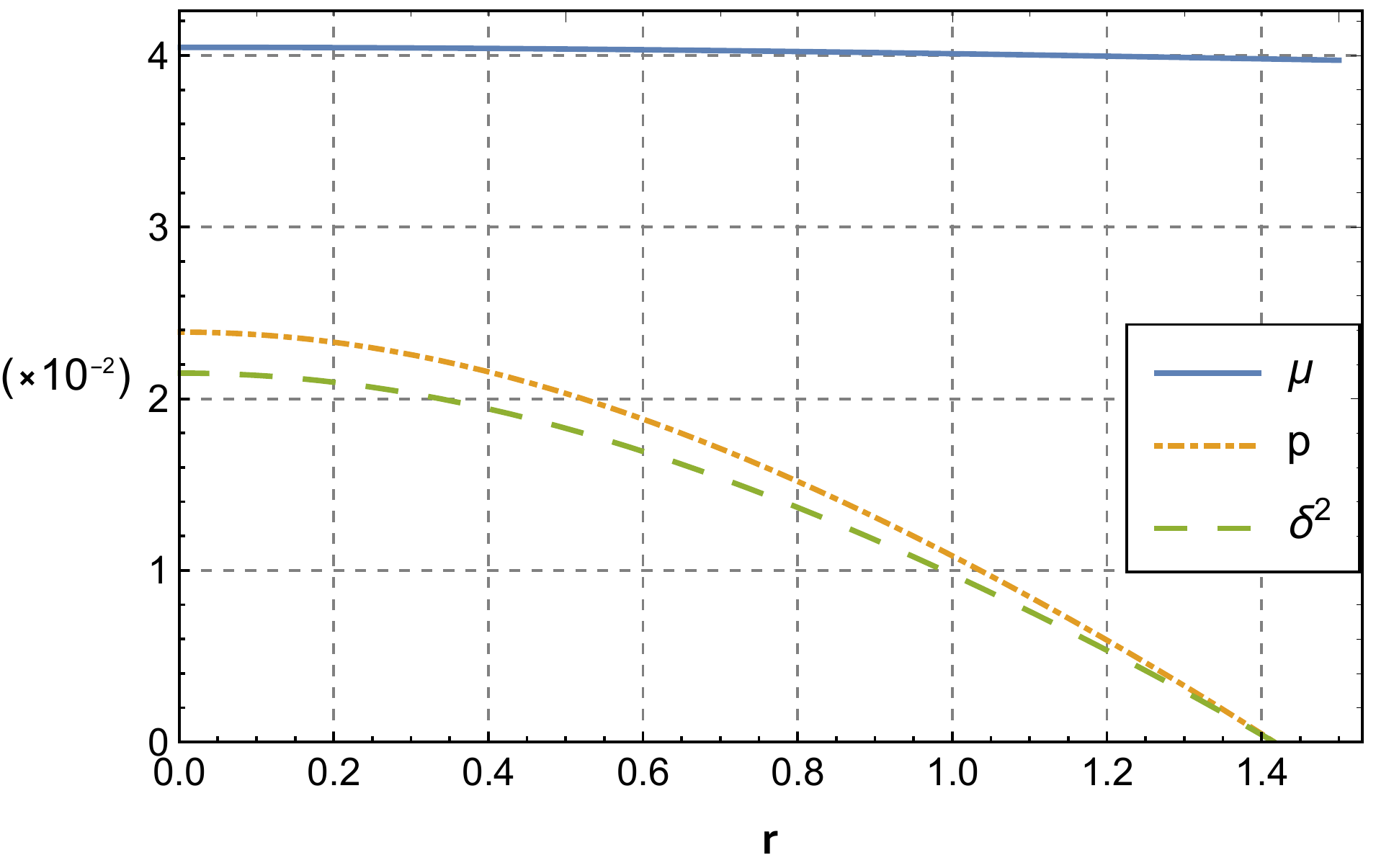}

}\caption{\label{Fig:Rec3}Plots of the behavior of the metric components, (a)
and matter variables (b) associated with the solution in Eqs.~\eqref{eq:Rec2_line_element_sph_sym_general},
\eqref{eq:Rec3_line_element_sph_sym_coef} and \eqref{eq:Rec2_matter_variables}
in the case $A_{0}=1$, $a=-3$, $b=1$, $c=3$, $d=0.03$, $r_{0}=1$
and $\gamma=0.9/(8\pi)$.}
\end{figure}

Another example,based on the same assumptions, can be given considering
\begin{equation}
A=A_{0}\left(a+\sqrt{c-be^{\rho}}\right)^{2}~,
\end{equation}
which corresponds to 
\begin{equation}
Y=-\frac{be^{\rho}}{2\sqrt{c-be^{\rho}}\left(a+\sqrt{c-be^{\rho}}\right)}~,
\end{equation}
Eq.~\eqref{eq:Rec2_constr_DgM} then gives 
\begin{equation}
\mathcal{K}=\frac{c\psi\left(a\sqrt{c-be^{\rho}}-2be^{\rho}+c\right)}{\left(c-be^{\rho}\right)\left[4\psi\left(a\sqrt{c-be^{\rho}}-2be^{\rho}+c\right)-b\,de^{\rho}\right]}\,,
\end{equation}
with 
\begin{equation}
\psi=\left(\frac{\sqrt{a^{2}+8c}+a+4\sqrt{c-be^{\rho}}}{\sqrt{a^{2}+8c}-a-4\sqrt{c-be^{\rho}}}\right)^{\frac{a}{\sqrt{a^{2}+8c}}}\,.
\end{equation}
Using the area radius $r$, we find the following solution for the
metric \eqref{eq:Rec2_line_element_sph_sym_general}

\begin{equation}
\begin{aligned}A & =A_{0}\left[a+y\left(r\right)\right]^{2}\,,\\
B & =\frac{4c\left[ay\left(r\right)+2y\left(r\right)^{2}-c\right]}{y\left(r\right)^{2}\left[4ay\left(r\right)+8y\left(r\right)^{2}+d\,\psi\left(r\right)\left(y\left(r\right)^{2}-c\right)-4c\right]}\,,
\end{aligned}
\label{eq:Rec3_line_element_sph_sym_coef}
\end{equation}
where 
\begin{equation}
\begin{aligned}y\left(r\right) & =\sqrt{c-\frac{br^{2}}{r_{0}^{2}}}\,,\\
\psi\left(r\right) & =\left(\frac{\sqrt{a^{2}+8c}+a+4y\left(r\right)}{\sqrt{a^{2}+8c}-a-4y\left(r\right)}\right)^{\frac{a}{\sqrt{a^{2}+8c}}}\,,
\end{aligned}
\end{equation}
with the following expressions for the energy density and pressure
of the fluid

\begin{equation}
\begin{aligned}\mu & =\frac{b\,d\,\gamma\left[6y^{5}+7ay^{4}+2y^{3}\left(a^{2}-3c\right)-4acy^{2}\right]\psi}{16\pi c(\gamma-1)r_{0}^{2}(a+y)[c-y(a+2y)]^{2}}+\\
 & \hspace{0.4cm}+\frac{b\,d\,\gamma c^{2}\left(2y+a\right)\psi}{16\pi c(\gamma-1)r_{0}^{2}(a+y)[c-y(a+2y)]^{2}}-\\
 & \hspace{0.4cm}-\frac{bd\left[6y^{4}+3ay^{3}-5cy^{2}+2c^{2}\right]\psi}{32\pi c(\gamma-1)r_{0}^{2}[c-y(a+2y)]^{2}}+\\
 & \hspace{0.4cm}+\frac{b[2\gamma(2a+3y)-3(a+y)]}{8\pi cr_{0}^{2}(\gamma-1)(a+y)}\,,\\
p & =\frac{b\,d\,y\left(ay-2c+3y^{2}\right)\psi}{32\pi c(\gamma-1)r_{0}^{2}(a+y)\left(ay-c+2y^{2}\right)}+\\
 & \hspace{0.4cm}+\frac{b(a+3y)}{8\pi cr_{0}^{2}(\gamma-1)(a+y)}\,,
\end{aligned}
\label{eq:Rec3_matter_variables}
\end{equation}
In Figure \ref{Fig:Rec3} we show the behavior of this solution for
specific values of the parameters. Notice that also this solution
admits the existence of a common hypersurface where both $p$ and
$\delta$ vanish.

Before finishing this section we remark that, as shown by Figs. \ref{Fig:Rec1_1} - \ref{Fig:Rec3}, in all considered cases it is possible to find values of the parameters for which all the thermodynamical quantities and spin density are positive, hence, all the classical energy conditions are valid.

\section{Generating theorems\label{sec:Generating_theorems}}

As discussed in Ref.~\citep{Sante1}, the form of the structure equations
\eqref{eq:TOV_static_P} - \eqref{eq:TOV_static_D1} is especially
useful to find algorithms for generating new solutions from previous
known ones.

Consider a solution for the structure equations \eqref{eq:TOV_static_P}
- \eqref{eq:TOV_static_D1} characterized by the functions 
\begin{equation}
\left\{ \mathcal{P}_{0},\mathcal{M}_{0},\Delta_{0},K_{0},\mathcal{E}_{0},\mathbb{X}_{0},\mathbb{Y}_{0},\left(\mathbb{B}_{1}\right)_{0},\left(\mathbb{B}_{2}\right)_{0},\left(\mathbb{D}_{1}\right)_{0},\left(\mathbb{D}_{2}\right)_{0}\right\} \,.\label{eq:Generating_theorems_Linear_original_solution_functions}
\end{equation}
Given the quantities 
\begin{equation}
\begin{aligned}\mathcal{M} & =\mathcal{M}_{0}+\mathcal{M}_{1}\,,\\
\mathcal{P} & =\mathcal{P}_{0}+\mathcal{P}_{1}\,,\\
\Delta^{2} & =\Delta_{0}{}^{2}+\Delta_{1}{}^{2}\,,\\
\mathcal{K} & =\mathcal{K}_{0}+\mathcal{K}_{1}\,,
\end{aligned}
\label{eq:Generating_theorems_Linear_deformed_quantities_definition}
\end{equation}
where $\left\{ \mathcal{P}_{1},\mathcal{M}_{1},\Delta_{1},\mathcal{K}_{1}\right\} $
are sufficiently smooth arbitrary functions, let us search conditions
on the deforming functions to so that the set $\left\{ \mathcal{P},\mathcal{M},\Delta,\mathcal{K},\mathcal{E},\mathbb{X},\mathbb{Y},\mathbb{B}_{1},\mathbb{B}_{2},\mathbb{D}_{1},\mathbb{D}_{2}\right\} $
is a solution of the structure equations.

Substituting Eq.~\eqref{eq:Generating_theorems_Linear_deformed_quantities_definition}
in Eq.~\eqref{eq:TOV_static_K} we find
\begin{align}
\partial_{\rho}\mathcal{K}_{1}+2\mathcal{K}_{1}{}^{2}+2\mathcal{K}_{0}\left(\Delta_{1}{}^{2}-\mathcal{M}_{1}\right)-\nonumber \\
-\mathcal{K}_{1}\left(2\mathcal{M}_{0}-4\mathcal{K}_{0}-2\Delta_{0}{}^{2}+2\Delta_{1}{}^{2}-2\mathcal{M}_{1}+\frac{1}{2}\right) & =0\,.\label{eq:Generating_theorems_Linear_riccati_K}
\end{align}
This equation has the form of a Riccati differential equation to which,
in general, there are no known closed form solutions. We can, nonetheless,
consider particular cases so that the previous equation reduces to
a Bernoulli differential equation, where general closed form solutions
exist.

\subsection{Case 1\label{subsec:Generating_theorems_Case_1}}

Let us first consider that 
\begin{equation}
\mathcal{M}_{1}=\Delta_{1}^{2}\,.\label{eq:Generating_theorems_Linear_case_1_M1}
\end{equation}
In this case, Eq.~\eqref{eq:Generating_theorems_Linear_riccati_K}
can be readily integrated for $\mathcal{K}_{1}$, such that 
\begin{equation}
\mathcal{K}_{1}=0\:\lor\:\mathcal{K}_{1}\left(\rho\right)=\frac{\text{Exp}\left[\int_{\rho_{0}}^{\rho}\Lambda\,dx\right]}{\mathcal{K}_{\star}+2\int_{\rho_{0}}^{\rho}\text{Exp}\left[\int_{y_{0}}^{y}\Lambda\,dx\right]\,dy}\,,\label{eq:Generating_theorems_Linear_case_1_K1_solution}
\end{equation}
where $\mathcal{K}_{\star}$ is an integration constant and 
\begin{equation}
\Lambda=2\mathcal{M}_{0}-2\text{\ensuremath{\Delta}}_{0}{}^{2}-4\ensuremath{\mathcal{K}_{0}}+\frac{1}{2}\,.
\end{equation}
Using Eq.~\eqref{eq:Generating_theorems_Linear_case_1_M1} in Eq.~\eqref{eq:TOV_static_P}
we find

\begin{align}
\partial_{\rho}\mathcal{P}_{1}+\mathcal{P}_{1}\left(3\mathcal{K}_{0}+3\mathcal{K}_{1}-\mathcal{M}_{0}+2\mathcal{P}_{0}-\Delta_{0}{}^{2}-\frac{7}{4}\right)+\nonumber \\
+\mathcal{P}_{1}{}^{2}+\frac{1}{4}\mathcal{F} & =0\,,\label{eq:Generating_theorems_Linear_case_1_P1}
\end{align}
with
\begin{equation}
\begin{aligned}\mathcal{F} & =4\mathcal{M}_{0}\left(\Delta_{1}{}^{2}+\mathcal{K}_{1}\right)-8\left(\mathcal{P}_{0}+\mathcal{P}_{1}\right)\Delta_{1}{}^{2}+12\mathcal{P}_{0}\mathcal{K}_{1}+\\
 & +4\Delta_{0}{}^{2}\Delta_{1}{}^{2}-4\partial_{\rho}\Delta_{1}{}^{2}+4\Delta_{1}{}^{4}+7\Delta_{1}{}^{2}-\\
 & -12\Delta_{1}{}^{2}\mathcal{K}_{0}-16\Delta_{0}{}^{2}\mathcal{K}_{1}-12\Delta_{1}{}^{2}\mathcal{K}_{1}\,.
\end{aligned}
\end{equation}
For Eq.~\eqref{eq:Generating_theorems_Linear_case_1_P1} to reduce
to a Bernoulli like differential equation we will require $\mathcal{F}\left(\rho\right)=0$,
that is 
\begin{align}
\partial_{\rho}\Delta_{1}{}^{2}-\Delta_{1}{}^{4}-\mathcal{K}_{1}\left(\mathcal{M}_{0}+3\mathcal{P}_{0}-4\Delta_{0}{}^{2}\right)-\nonumber \\
-\Delta_{1}{}^{2}\left(\frac{7}{4}+\mathcal{M}_{0}-3\mathcal{K}_{0}-3\mathcal{K}_{1}-2\mathcal{P}_{0}-2\mathcal{P}_{1}+\Delta_{0}{}^{2}\right) & =0\,,\label{eq:Generating_theorems_Linear_case_1_F1_zero}
\end{align}
which, by setting $\mathcal{K}_{1}=0$ or $\mathcal{M}_{0}+3\mathcal{P}_{0}-4\Delta_{0}{}^{2}=0$,
can be formally solved, such that
\begin{equation}
\Delta_{1}^{2}\left(\rho\right)=0\:\lor\:\Delta_{1}{}^{2}\left(\rho\right)=\frac{\text{Exp}\left[\int_{\rho_{0}}^{\rho}\Phi\,dx\right]}{\Delta_{\star}-\int_{\rho_{0}}^{\rho}\text{Exp}\left[\int_{y_{0}}^{y}\Phi\,dx\right]\,dy}\,,\label{eq:Generating_theorems_Linear_case_1_Delta1_solution}
\end{equation}
where $\Delta_{\star}$ is an integration constant and 
\begin{equation}
\Phi=\mathcal{M}_{0}-3\mathcal{K}_{0}-3\mathcal{K}_{1}-2\mathcal{P}_{0}-2\mathcal{P}_{1}+\Delta_{0}{}^{2}+\frac{7}{4}\,.
\end{equation}
Consequently, from Eq.~\eqref{eq:Generating_theorems_Linear_case_1_P1},
we find 
\begin{equation}
\mathcal{P}_{1}\left(\rho\right)=0\:\lor\:\mathcal{P}_{1}\left(\rho\right)=\frac{\text{Exp}\left[\int_{\rho_{0}}^{\rho}\Gamma\,dx\right]}{\mathcal{P}_{\star}+\int_{\rho_{0}}^{\rho}\text{Exp}\left[\int_{y_{0}}^{y}\Gamma\,dx\right]\,dy}\,,\label{eq:Generating_theorems_Linear_case_1_P1_solution}
\end{equation}
with
\begin{equation}
\Gamma=\mathcal{M}_{0}-2\mathcal{P}_{0}+\Delta_{0}{}^{2}-3\mathcal{K}_{0}-3\mathcal{K}_{1}+\frac{7}{4}\,,
\end{equation}
and $\mathcal{P}_{\star}$ is an integration constant.

Before we conclude this Section, we should stress that Eqs.~\eqref{eq:Generating_theorems_Linear_case_1_K1_solution},
\eqref{eq:Generating_theorems_Linear_case_1_Delta1_solution} and
\eqref{eq:Generating_theorems_Linear_case_1_P1_solution} present
two possible solutions for the considered functions and all combinations
of those solutions verify the structure equations with $\mathcal{M}_{1}=\Delta_{1}^{2}$,
leading, \emph{a priori}, to distinct solutions.

\subsection{Case 2}

Another possibility to solve Eq.~\eqref{eq:Generating_theorems_Linear_riccati_K}
is the case when 
\begin{equation}
\left(\mathcal{K}_{0}+\mathcal{K}_{1}\right)\left(2\Delta_{1}^{2}-2\mathcal{M}_{1}\right)=G\left(\rho\right)\mathcal{K}_{1}+Q\left(\rho\right)\mathcal{K}_{1}{}^{2}\,,\label{eq:Generating_theorems_Linear_case_2_Original_assumption}
\end{equation}
where $G\left(\rho\right)$ and $Q\left(\rho\right)$ are sufficiently
smooth, arbitrary functions. Setting
\begin{equation}
\begin{aligned}2\Delta_{1}^{2}-2\mathcal{M}_{1} & =\mathcal{K}_{1}Q\left(\rho\right)\,,\\
G\left(\rho\right) & =\mathcal{K}_{0}Q\left(\rho\right)\,,
\end{aligned}
\label{eq:Generating_theorems_Linear_case_2_OA_implies}
\end{equation}
and substituting Eqs.~\eqref{eq:Generating_theorems_Linear_case_2_Original_assumption}
and \eqref{eq:Generating_theorems_Linear_case_2_OA_implies} in Eq.~\eqref{eq:Generating_theorems_Linear_riccati_K}
we find
\begin{align}
\partial_{\rho}\mathcal{K}_{1}+\mathcal{K}_{1}\left[2\Delta_{0}{}^{2}-2\mathcal{M}_{0}+4\mathcal{K}_{0}+\mathcal{K}_{0}Q\left(\rho\right)-\frac{1}{2}\right]+\nonumber \\
+\left[2+Q\left(\rho\right)\right]\mathcal{K}_{1}{}^{2} & =0\,,\label{eq:Generating_theorems_Linear_case_2_K1_bernoulli_equation}
\end{align}
which, provided an expression for $Q\left(\rho\right)$ can be solved
for $\mathcal{K}_{1}$, or vice-versa.

Now, to solve the remaining equations for the functions $Q$, $\mathcal{P}_{1}$
and $\Delta_{1}$, we will consider that the original solution is
such that $\mathcal{M}_{0}=\mathcal{P}_{0}=\Delta_{0}=0$, that is,
the original space-time is described by a vacuum solution of the field
equations. From Eq.~\eqref{eq:TOV_static_P} we then find 
\begin{equation}
\partial_{\rho}\mathcal{P}_{1}+\mathcal{P}_{1}{}^{2}+\mathcal{P}_{1}\left[3\mathcal{K}_{0}+3\mathcal{K}_{1}-\frac{7}{4}\right]+\mathcal{J}\left(\rho\right)=0\,,\label{eq:Generating_theorems_Linear_case_2_P1_Riccati_equation}
\end{equation}
where 
\begin{equation}
\begin{aligned}\mathcal{J}\left(\rho\right) & =-\partial_{\rho}\Delta_{1}{}^{2}+\Delta_{1}{}^{4}+\frac{1}{2}Q\left(\rho\right)\mathcal{K}_{1}\left[\mathcal{P}_{1}-\mathcal{K}_{0}-\mathcal{K}_{1}+\frac{1}{4}\right]\\
 & +\Delta_{1}{}^{2}\left(-2\mathcal{P}_{1}-\frac{1}{2}Q(x)\mathcal{K}_{1}-3\mathcal{K}_{0}-3\mathcal{K}_{1}+\frac{7}{4}\right)\,.
\end{aligned}
\end{equation}
As before, to reduce Eq.~\eqref{eq:Generating_theorems_Linear_case_2_P1_Riccati_equation}
to a Bernoulli differential equation we will impose $\mathcal{J}\left(\rho\right)=0$.
Unfortunately, this equation itself is also not possible to solve
in general since it has the form of a Riccati differential equation.
Let us then further impose the last term in the first line of the previous equation
to be zero. Solving for $\mathcal{K}_{1}$, we have 
\begin{equation}
\mathcal{K}_{1}=\mathcal{P}_{1}-\mathcal{K}_{0}+\frac{1}{4}\,,\label{eq:Generating_theorems_Linear_case_2_K1_solution}
\end{equation}
where we have ignored the solutions where $Q=0\,\lor\,\mathcal{K}_{1}=0$
since they lead to a particular case of subsection \ref{subsec:Generating_theorems_Case_1}.

Considering the constraint that originally we have a vacuum solution,
substituting Eq.~\eqref{eq:Generating_theorems_Linear_case_2_K1_solution}
in Eq.~\eqref{eq:Generating_theorems_Linear_case_2_K1_bernoulli_equation}
we find, 
\begin{equation}
Q=-\frac{8\left(2\partial_{\rho}\mathcal{P}_{1}+4\mathcal{P}_{1}{}^{2}+\mathcal{P}_{1}\right)}{\left(4\mathcal{P}_{1}+1\right)\left(4\mathcal{P}_{1}-4\mathcal{K}_{0}+1\right)}\,.\label{eq:Generating_theorems_Linear_case_6_Q_solution}
\end{equation}

Gathering the previous results we find the following expressions for
the remaining perturbations 
\begin{equation}
\begin{aligned}\mathcal{P}_{1} & =\frac{e^{\rho}}{\mathcal{P}_{\star}+4e^{\rho}}\,,\\
\Delta_{1}^{2}=0\:\lor\:\Delta_{1}{}^{2} & =\frac{\text{Exp}\left[-\int_{\rho_{0}}^{\rho}\Phi\,dx\right]}{\Delta_{\star}-\int_{\rho_{0}}^{\rho}\text{Exp}\left[-\int_{y_{0}}^{y}\Phi\,dx\right]\,dy}\,,\\
\mathcal{M}_{1} & =\Delta_{1}^{2}+\frac{2\partial_{\rho}\mathcal{P}_{1}+4\mathcal{P}_{1}{}^{2}+\mathcal{P}_{1}}{4\mathcal{P}_{1}+1}\,,
\end{aligned}
\end{equation}
where $\mathcal{P}_{\star}$ and $\Delta_{\star}$ are integrating
constants and 
\begin{equation}
\Phi=2\mathcal{P}_{1}+\frac{1}{2}Q(x)\mathcal{K}_{1}+3\mathcal{K}_{0}+3\mathcal{K}_{1}-\frac{7}{4}\,.
\end{equation}
Notice that we did not consider the case when $\mathcal{P}_{1}=0$
since it would lead to the case when $Q\left(\rho\right)=0$, which,
as mentioned before, represents a particular case of subsection \ref{subsec:Generating_theorems_Case_1}.
Let us also remark that, for solutions generated using the above equations,
the functional form of the pressure, $\mathcal{P}\equiv\mathcal{P}_{1}$,
is independent of the original solution and completely determined
up to a constant. Moreover, notice that the pressure - in such solutions
- is only null when $\rho\to-\infty$.

\subsection{Case 3}

Let us now consider the deformations in Eq.~\eqref{eq:Generating_theorems_Linear_deformed_quantities_definition}
with the extra constraint 
\begin{equation}
\begin{aligned}\mathbb{Y} & =\mathbb{Y}_{0}\,,\end{aligned}
\label{eq:Generating_theorems_Visser_constraint_Y}
\end{equation}
that is, we will impose that the function $\mathbb{Y}$ is unchanged
between the original and the perturbed space-time. This is a generalization
of the deformations considered in Refs.~\citep{Sante1,Boonserm1},
for non-null spin density. Substituting Eqs.~\eqref{eq:Generating_theorems_Linear_deformed_quantities_definition}
and \eqref{eq:Generating_theorems_Visser_constraint_Y} in Eq.~\eqref{eq:TOV_static_Y}
we find that 
\begin{equation}
\mathcal{P}_{1}=\Delta_{1}^{2}-\mathcal{K}_{1}\,.\label{eq:Generating_theorems_Visser_constraint_P1}
\end{equation}
Using Eqs.~\eqref{eq:Generating_theorems_Linear_deformed_quantities_definition},
\eqref{eq:Generating_theorems_Visser_constraint_Y} and \eqref{eq:Generating_theorems_Visser_constraint_P1}
in Eqs.~\eqref{eq:TOV_static_K} and \eqref{eq:TOV_static_P} we
find the following relations for $\mathcal{M}_{1}$ and $\mathcal{K}_{1}$

\begin{align}
\mathcal{M}_{1} & =\frac{\mathcal{K}_{1}\left(2\mathbb{Y}_{0}+3\right)}{2\mathbb{Y}_{0}+1}+\Delta_{1}^{2}\,,\label{eq:Generating_theorems_Visser_M1}\\
\mathcal{K}_{1} & =\frac{\text{Exp}\left[-\int_{\rho_{0}}^{\rho}\Phi\,dx\right]}{\mathcal{K}_{\star}-\int_{\rho_{0}}^{\rho}\frac{4}{2\mathbb{Y}_{0}+1}\text{Exp}\left[-\int_{y_{0}}^{y}\Phi\,dx\right]\,dy}-\mathcal{K}_{0}\,,\label{eq:Generating_theorems_Visser_K1}
\end{align}
where 
\begin{align}
\Phi & =\frac{\mathcal{K}_{0}\left(6+4\mathbb{Y}_{0}\right)}{2\mathbb{Y}_{0}+1}+2\Delta_{0}^{2}-2\mathcal{M}_{0}-\frac{1}{2}\,,\label{eq:Generating_theorems_Visser_Phi}
\end{align}
and $\mathcal{K}_{\star}$ is an integration constant. Eqs.~\eqref{eq:Generating_theorems_Visser_constraint_P1}
- \eqref{eq:Generating_theorems_Visser_Phi} generalize the results
in Ref.~\citep{Sante1} in the presence of a non-null spin density
\footnote{Notice that there is an error in the expression for $\mathcal{M}_{1}$
in Ref.~\citep{Sante1}. The correct expression is found by setting
$\Delta_{0}^{2}=\Delta_{1}^{2}=0$ in Eq.~\eqref{eq:Generating_theorems_Visser_M1}.}.

Contrary to the previous cases, the Eqs.~\eqref{eq:Generating_theorems_Visser_constraint_P1}
- \eqref{eq:Generating_theorems_Visser_Phi} do not completely determine
the system since the function $\Delta_{1}^{2}$ is unconstrained.
Notice that $\mathcal{K}\equiv\mathcal{K}_{1}$ is determined uniquely
by the unperturbed solution and $\Delta_{1}^{2}$ will only affect
$\mathcal{M}_{1}$ and $\mathcal{P}_{1}$. Therefore, provided an
unperturbed solution, the metric of the perturbed space-time is completely
determined by Eqs.~\eqref{eq:Generating_theorems_Visser_constraint_Y}
and \eqref{eq:Generating_theorems_Visser_K1}. As already pointed
out, $\Delta_{1}^{2}$ will not only affect the energy density and
the pressure of the fluid but also the Weyl tensor components. Thus,
although the metric of the space-time is independent of $\Delta_{1}^{2}$,
the geometry is profoundly influenced by the presence of spin.

\section{Conclusions\label{sec:Conclusions}}

In this paper we have used the 1+1+2 formalism to derive the structure
equations for LRS I and LRS II, stationary space-times with a Weyssenhoff
like torsion field in the context the ECSK theory of gravity. The
structure of the covariant equations show in detail how the spin interacts
with the space-time via the torsion tensor. In particular, the presence
of a torsion tensor field separates the magnetic part of the Weyl
tensor in two distinct tensors, which behave differently. Even in
the case of static LRS II space-times, the magnetic parts of the Weyl
tensor do not vanish and some of its components depend on both the
value and spatial derivative of the spin density. This suggests, in
particular, that the effects of spin on the matter fluid, even in
the regimes expected to be found in neutron stars, may not be negligible,
as it was previously thought  (see e.g. \citep{DemPros}), even in the case in which the contribution to the spin is very small.

The 1+1+2 equations were then used to derive the covariant Tolman-Oppenheimer-Volkoff
equations for ECSK gravity for LRSI and LRSII space-times. In the
case of LRSII space-times, the equations are structurally very similar
to the ones of GR. Indeed this similarity allows to recast them into
the same form of the GR TOV equations via a redefinition of the matter
variables and the electric part of the Weyl tensor. As a consequence
we found that at the level of the metric it is possible to map static,
locally rotationally symmetric solutions of class II from the Einstein-Cartan
theory to the ones of the theory of General Relativity. Moreover,
due to this mapping and the re-scale in the matter variables, some
GR solutions which are physically irrelevant become, in the context
of ECSK gravity, interesting.

When we examine in detail the physical properties of physically relevant
solutions, the differences between the the Einstein-Cartan theory
and GR become once more evident. This is particularly true looking
at junction conditions. We found that the requirement that all of
the components of the Riemann tensor have finite discontinuities across
the separation surface $\mathcal{N}$ leads to additional constraints
with respect to the tornsionless case. This is especially evident
when looking at the structure equations for stationary LRSI and LRS
II space-times sourced by a Weyssenhoff fluid. In these equations
the magnetic parts of the Weyl tensor depend explicitly on the derivatives
of the torsion tensor and the classical Israel junction conditions
of GR do not guarantee these terms to be finite across $\mathcal{N}$.
As consequence of the generalized junction conditions, in the considered
setup, observers at the interior and exterior space-times must measure
the same value for the spin density at $\mathcal{N}$, turning the
task of finding physically relevant solutions even more daunting.

Using the full set of structure equations and boundary conditions
provided by the junction formalism, we were able to study various
properties of possible solutions. We started by analyzing how the
presence of spin changes the Buchdahl limit for the maximum compactness
of a star. We concluded that the spin-geometry coupling allows stars
with a given circumferential radius to hold more matter than the corresponding
GR ones. Next we considered the case of static, spherically symmetric
compact objects entirely held by the matter spin, smoothly matched
to a vacuum exterior. This scenario was expected to represent a good
model for cold neutron stars, where the thermodynamical pressure is
negligible when compared to the spin density. We found, surprisingly,
that such objects cannot be simultaneously static, spherically symmetric
and smoothly matched to a vacuum exterior. This is a strong result
and it is necessary to discuss in detail the hypothesis that led to
such conclusion. More specifically our conclusion may not be valid
if:
\begin{enumerate}[label=\roman*.]
\item the spin density is not a monotonically decreasing function of the
radial coordinate inside the star; 
\item we consider a non-vacuum exterior space-time; 
\item we replace the uncharged Weyssenhoff fluid model; 
\item we allow the presence of a thin shell. 
\end{enumerate}
The first possibility might lead to a total energy density and a pressure
density which is not monotonically decreasing. While this is not a
strong enough reason to discard this case, we expect these oscillation
to make the solution unstable under small perturbations. The second
case suggests that if ECSK theory had a non trivial vacuum (vortical)
solution, one could smoothly match the interior to it, bypassing the
requirement of the spin density to vanish at an hypersurface. At present
there is no evidence that such solution might/should exist. Indeed
the theory is expected to reduce to GR in vacuum. For what concerns
hypothesis (iii), the Weyssenhoff fluid can be advocated to be a good
model for the matter fluids that might constitute cold neutron stars.
However, in this work we made the simplifying assumption that the
fluid is electrically neutral. If instead a charged Weyssenhoff fluid
model is considered, we expect that other effects will appear - such
as anisotropic pressure - which may drastically change the behavior
of the fluid. As for the last possibility, although a smooth junction
with a vacuum exterior might represent a more reasonable scenario,
it might be argued that neutron stars may have a well defined surface,
therefore it is not completely unreasonable to consider the presence
of a thin shell of matter at the matching surface.

On top of the zero pressure solution considered above, we have also
considered solutions in which pressure is non zero. Using reconstruction
algorithms we have been able to obtain various classes of solutions
for the interior of static, spherically symmetric compact objects
that can be smoothly matched to a Schwarzschild exterior. One family
of those solutions, which we dubbed Buchdahl stars, represent a very
interesting scenario: they admit the existence of a common hypersurface
where the pressure, spin density and energy density all vanish. This
models, studied for the first time by Buchdahl for gaseous stars in
GR \citep{Buch3}, represents the scenario where the fluid that composes
a star will smoothly dissipate away from a denser core and transition
to vacuum. These solutions also provided a key example for the effects
that spin may have on the behavior of the fluid. Figures.~\ref{Fig:Rec1_1}
- \ref{Fig:Rec1_3} clearly exemplify that even if the spin density
is much smaller than the other matter variables, it allows for a much
richer behavior for the fluid.

The natural question that emerges is about the stability of these solutions. 
Because of the non trivial role of the magnetic part of the Weyl tensor, no standard ``zeroth order'' (and Newtonian based) criterion is necessarily valid in our case. In cases in which the spin density is decreasing, one very heuristic criterium of stability of our solutions is to guarantee (as we have done) that both energy density and pressure of the fluid are decreasing functions of the radial coordinate when the spin is small at least in a non empty set of values of the parameters. However, a complete study of the stability of the solutions we have found requires a more careful study, which will be the topic of a series of future works. 

Finally, as in the case of GR, also in ECSK theory it is possible
to derive generating theorems. In this work we have presented several
algorithms to generate new exact solutions from previously known ones.
We should stress here that the results we obtained followed from the
simple idea of finding conditions so that the Riccati differential
equations would reduce to Bernoulli equations. Although this scheme
allowed us to find various generating algorithms, we make no claim
that we have exhausted all possibilities for finding new ones. On
this note, the integrability conditions for Riccati type equations
in Refs.~\citep{Ramos,Mak,Lobo} were also considered. However, these
did not lead useful results in the considered context.
\begin{acknowledgments}
PL is grateful to IDPASC and FCT-Portugal for financial support through
Grant No. PD/BD/114074/2015. SC is supported by the Funda\c{c}\~{a}o
para a Ci\^{e}ncia e Tecnologia through project IF/00250/2013 and
acknowledge financial support provided under the European Union's
H2020 ERC Consolidator Grant ``Matter and strong-field gravity: New
frontiers in Einstein's theory'' grant agreement No. MaGRaTh646597. 
We also thank the anonymous referee for useful suggestions.
\end{acknowledgments}

\appendix

\section{Covariantly defined quantities for the derivatives of the tangent
vectors\label{Appendix: Covariant_quantities}}

Using the definitions of the projector operators onto the hypersurfaces
$V$ and $W$, let us show how the covariant derivatives of the tangent
vector fields $v$ and $e$ can be uniquely decomposed in their components
along $u$, $e$ and $W$.

\subsection{Decomposition on the sheet $W$}

Let us first consider the projected covariant derivatives of the tensors
$u$ and $e$ on the sheet. These can be uniquely decomposed as

\begin{equation}
\delta_{\alpha}u_{\beta}\equiv N_{\alpha}{}^{\sigma}N_{\beta}{}^{\gamma}\nabla_{\sigma}u_{\gamma}=\frac{1}{2}N_{\alpha\beta}\tilde{\theta}+\Sigma_{\alpha\beta}+\varepsilon_{\alpha\beta}\Omega\,,
\end{equation}
where 
\begin{equation}
\begin{aligned}\tilde{\theta} & =\delta_{\alpha}u^{\alpha}\,,\\
\Sigma_{\alpha\beta} & =\delta_{\{\alpha}u_{\beta\}}\,,\\
\Omega & =\frac{1}{2}\varepsilon^{\sigma\gamma}\delta_{\sigma}u_{\gamma}\,,
\end{aligned}
\label{eqA:W_projected_cov_u_quantities}
\end{equation}
and 
\begin{equation}
\delta_{\alpha}e_{\beta}=\frac{1}{2}N_{\alpha\beta}\phi+\zeta_{\alpha\beta}+\varepsilon_{\alpha\beta}\xi\,,\label{eqA:W_projected_cov_space-like_e}
\end{equation}
with 
\begin{equation}
\begin{aligned}\phi & =\delta_{\alpha}e^{\alpha}\,,\\
\zeta_{\alpha\beta} & =\delta_{\{\alpha}e_{\beta\}}\,,\\
\xi & =\frac{1}{2}\varepsilon^{\sigma\gamma}\delta_{\sigma}e_{\gamma}\,,
\end{aligned}
\label{eqA:W_projected_cov_e_quantities}
\end{equation}
where the curly brackets represent the projected symmetric part without
trace of a tensor in $W$, that is, for a tensor $\psi_{\alpha\beta}$,
\begin{equation}
\psi_{\left\{ \alpha\beta\right\} }=\left[N^{\mu}{}_{(\alpha}N_{\beta)}\,^{\nu}-\frac{N_{\alpha\beta}}{2}N^{\mu\nu}\right]\psi_{\mu\nu}\,.\label{eqA:curly_notation_definition}
\end{equation}
Using the 2-form volume $\varepsilon_{\alpha\beta}$ a completely
anti-symmetric tensor defined on the sheet, $\psi_{\left[\alpha\beta\right]}$,
can be written as 
\begin{equation}
\psi_{\left[\alpha\beta\right]}=\varepsilon_{\alpha\beta}\left(\frac{1}{2}\epsilon^{\gamma\sigma}\psi_{\gamma\sigma}\right)\,.
\end{equation}
This property was used in Eq.~\eqref{eq:volume_forms}.

\subsection{Decomposition on $V$}

The decomposition of the projected covariant derivatives of $u^{\alpha}$
onto $V$, is given by 
\begin{equation}
D_{\alpha}u_{\beta}=h_{\alpha}^{\sigma}h_{\beta}^{\gamma}\nabla_{\sigma}u_{\gamma}=\frac{1}{3}h_{\alpha\beta}\theta+\sigma_{\alpha\beta}+\omega_{\alpha\beta}\,,\label{eqA:H_decomposition_cov_u}
\end{equation}
with 
\begin{align}
\theta & =h^{\alpha\beta}D_{\alpha}u_{\beta}\,,\label{eqA:H_decomposition_cov_expansion}\\
\sigma_{\alpha\beta} & =D_{\left\langle \alpha\right.}u_{\left.\beta\right\rangle }\,,\\
\omega_{\alpha\beta} & =h^{\sigma}{}_{[\alpha}h_{\beta]}{}^{\gamma}D_{\sigma}u_{\gamma}\,,\label{eqA:H_decomposition_cov_vorticity}
\end{align}
where we used the angular brackets to represent the projected symmetric
part without trace of a tensor on $V$, that is, for a tensor, $\psi_{\alpha\beta}$,
\begin{equation}
\psi_{\left\langle \alpha\beta\right\rangle }=\left[h^{\mu}{}_{(\alpha}h_{\beta)}{}^{\nu}-\frac{h_{\alpha\beta}}{3}h^{\mu\nu}\right]\psi_{\mu\nu}\,.\label{eqA:angular_brackets_definition}
\end{equation}
The scalar and tensor quantities in Eqs.~\eqref{eqA:H_decomposition_cov_expansion}
- \eqref{eqA:H_decomposition_cov_vorticity} can themselves be further
decomposed in their contributions exclusively on $W$ and along $e$,
such that 
\begin{equation}
\theta=\tilde{\theta}+\bar{\theta}\,,
\end{equation}
where $\tilde{\theta}$ is defined in Eq.~\eqref{eqA:W_projected_cov_u_quantities}
and 
\begin{equation}
\bar{\theta}=-u_{\beta}\left(e^{\alpha}D_{\alpha}e^{\beta}\right)=-u_{\beta}\hat{e}^{\beta}\,;
\end{equation}
\begin{equation}
\sigma_{\alpha\beta}=\Sigma_{\alpha\beta}+2\Sigma_{(\alpha}e_{\beta)}+\Sigma\left(e_{\alpha}e_{\beta}-\frac{1}{2}N_{\alpha\beta}\right)\,,
\end{equation}
with 
\begin{equation}
\begin{aligned}\Sigma_{\alpha\beta} & =\sigma_{\left\{ \alpha\beta\right\} }\,,\\
\Sigma_{\alpha} & =N_{\alpha}^{\gamma}e^{\beta}\sigma_{\gamma\beta}\,,\\
\Sigma & =e^{\alpha}e^{\beta}\sigma_{\alpha\beta}=-N^{\alpha\beta}\sigma_{\alpha\beta}\,,
\end{aligned}
\end{equation}
and 
\begin{equation}
\omega_{\alpha\beta}=\varepsilon_{\alpha\beta}\Omega-\varepsilon_{\alpha\lambda}\omega^{\lambda}e_{\beta}+e_{\alpha}\varepsilon_{\beta\lambda}\omega^{\lambda}\,,
\end{equation}
where $\Omega$ is given in Eq.~\eqref{eqA:W_projected_cov_u_quantities}
and 
\begin{equation}
\omega^{\lambda}=\frac{1}{2}\varepsilon^{\mu\nu\lambda}D_{\mu}u_{\nu}\,,
\end{equation}
which can be itself decomposed as 
\begin{equation}
\omega^{\lambda}=\Omega e^{\lambda}+\Omega^{\lambda}\quad,\text{with }\Omega^{\lambda}=N_{\alpha}^{\lambda}\omega^{\alpha}=\frac{1}{2}N_{\alpha}^{\lambda}\varepsilon^{\mu\nu\alpha}D_{\mu}u_{\nu}\,,
\end{equation}
therefore, equivalently, 
\begin{equation}
\omega_{\alpha\beta}=\varepsilon_{\alpha\beta\gamma}\left(\Omega e^{\gamma}+\Omega^{\gamma}\right)\,.
\end{equation}
The quantities $\theta$, $\Sigma$, $\tilde{\theta}$ and $\bar{\theta}$
are not independent, in fact: 
\begin{align}
\bar{\theta} & =\frac{1}{3}\theta+\Sigma\,,\\
\tilde{\theta} & =\frac{2}{3}\theta-\Sigma\,;
\end{align}
as such, when setting up the 1+1+2 formalism only two are chosen.
The convention followed here uses the variables $\theta$ and $\Sigma$.

For the projected covariant derivative of the vector field $e$ on
$V$ we have 
\begin{equation}
D_{\alpha}e_{\beta}=h_{\alpha}{}^{\sigma}h_{\beta}{}^{\gamma}\nabla_{\sigma}e_{\gamma}=\delta_{\alpha}e_{\beta}+e_{\alpha}a_{\beta}\,,
\end{equation}
where $\delta_{\alpha}e_{\beta}$ is given by Eq.~\eqref{eqA:W_projected_cov_space-like_e}
and 
\begin{equation}
a_{\alpha}=e^{\mu}D_{\mu}e_{\alpha}=\hat{e}_{\alpha}\,.
\end{equation}

\subsection{Decomposition on the full manifold}

Finally, we can decompose the total covariant derivatives of $u^{\alpha}$
and $e^{\alpha}$, such that 
\begin{equation}
\nabla_{\alpha}u_{\beta}=-u_{\alpha}\left(\mathcal{A}e_{\beta}+\mathcal{A}_{\beta}\right)+D_{\alpha}u_{\beta}\,,
\end{equation}
with 
\begin{equation}
\begin{aligned}\mathcal{A} & =-u_{\gamma}u^{\mu}\nabla_{\mu}e^{\gamma}=-u_{\gamma}\dot{e}^{\gamma}\,,\\
\mathcal{A}_{\alpha} & =N_{\alpha\beta}\dot{u}^{\beta}\,,
\end{aligned}
\label{eqA:u_acceleration_cov_quantities}
\end{equation}
and 
\begin{align}
\nabla_{\alpha}e_{\beta} & =D_{\alpha}e_{\beta}-u_{\alpha}\alpha_{\beta}-\mathcal{A}u_{\alpha}u_{\beta}+\left[\frac{1}{3}\theta+\Sigma\right]e_{\alpha}u_{\beta}+\nonumber \\
 & \hspace{0.4cm}+\left[\Sigma_{\alpha}-\varepsilon_{\alpha\sigma}\Omega^{\sigma}\right]u_{\beta}\,,
\end{align}
where 
\begin{equation}
\alpha_{\mu}=h_{\mu}^{\sigma}\dot{e}_{\sigma}\,.
\end{equation}

\subsection{The actual physical kinematical variables\label{Appendix:subsec_Physical_kinematical_quantities}}

As discussed in Refs.~\citep{Luz,Paoli,Liberati,Speziale}, the presence
of a generic torsion field will affect the definition of the kinematical
quantities that characterize a congruence of curves, such that, $\theta$,
$\sigma_{\alpha\beta}$ and $\omega_{\alpha\beta}$, Eqs.~\eqref{eqA:H_decomposition_cov_expansion}
- \eqref{eqA:H_decomposition_cov_vorticity}, in general, do not represent
the actual geometric - physical - expansion, shear and vorticity of
the time-like congruence to which $u$ is tangent. These, however,
are related with the actual kinematical quantities by 
\begin{equation}
\begin{aligned}\theta_{g} & =\theta+W_{\sigma}^{\sigma}\,,\\
\sigma_{g\,\alpha\beta} & =\sigma_{\alpha\beta}+W_{\left\langle \alpha\beta\right\rangle }\,,\\
\omega_{g\,\alpha\beta} & =\omega_{\alpha\beta}+W_{\left[\alpha\beta\right]}\,,
\end{aligned}
\label{eqA:kinematical_quantities_relations_1}
\end{equation}
where we have used the index\emph{ g }to represent the physical -
geometric - kinematical quantities and the definition of angular brackets
is given in Eq.~\eqref{eqA:angular_brackets_definition}. In the
same way the presence of the torsion field will modify the kinematical
quantities $\phi$, $\zeta_{\alpha\beta}$ and $\xi$, Eqs.~\eqref{eqA:W_projected_cov_e_quantities}.
In particular we have the following relations 
\begin{equation}
\begin{aligned}\phi_{g} & =\phi+2S_{\gamma\mu\nu}e^{\gamma}N^{\mu\nu}\,,\\
\zeta_{g\,\alpha\beta} & =\zeta_{\alpha\beta}+2S_{\gamma\mu\nu}e^{\gamma}N_{\left\langle \alpha\right|}^{\mu}N_{\left|\beta\right\rangle }^{\nu}\,,\\
\xi_{g} & =\xi+S_{\gamma\mu\nu}e^{\gamma}\varepsilon^{\mu\nu}\,.
\end{aligned}
\label{eqA:kinematical_quantities_relations_2}
\end{equation}
In the particular setup that we propose to study - Weyssenhoff like
torsion - the extra terms in the RHS of Eqs.~\eqref{eqA:kinematical_quantities_relations_1}
and \eqref{eqA:kinematical_quantities_relations_2} that depend explicitly
of the torsion tensor will be null, therefore, in our case, the indicated
quantities will correspond to the actual geometric kinematical quantities.

\end{document}